\documentclass[aps,preprintnumbers,amsmath,amssymb,twocolumn,longbibliography]{revtex4-1}

\usepackage{graphicx}
\usepackage{dcolumn}
\usepackage{bm}
\usepackage{color}
\usepackage{amssymb}   
\usepackage{amsthm}

\begin{document}

\title{Quasicrystal formation in binary soft matter mixtures}

\author{A. Scacchi$^1$}
\author{W. R. C. Somerville$^2$}
\author{D. M. A. Buzza$^3$}
\author{A. J. Archer$^1$}

\affiliation{1. Department of Mathematical Sciences and Interdisciplinary Centre for Mathematical Modelling, Loughborough University, Loughborough LE11 3TU, United Kingdom\\
2. The MacDiarmid Institute for Advanced Materials and Nanotechnology, School of Chemical and Physical Sciences,Victoria University of Wellington, PO Box 600 Wellington, New Zealand\\
3. Department of Physics and Mathematics, University of Hull, Hull HU6 7RX, United Kingdom}

\begin{abstract}
Using a strategy that may be applied in theory or in experiments, we identify the regime in which a model binary soft matter mixture forms quasicrystals. The system is described using classical density functional theory combined with integral equation theory. Quasicrystal formation requires particle ordering with two characteristic lengthscales in certain particular ratios. How the lengthscales are related to the form of the pair interactions is reasonably well understood for one component systems, but less is known for mixtures. In our model mixture of big and small colloids confined to an interface, the two lengthscales stem from the range of the interactions between pairs of big particles and from the cross big-small interactions, respectively. The small-small lengthscale is not significant. Our strategy for finding quasicrystals involves tuning locations of maxima in the dispersion relation, or equivalently in the liquid state partial static structure factors.
\end{abstract}
\maketitle

For systems of soft particles of a single type, there is growing understanding of the ingredients required for the self-assembly of quasicrystals (QCs). The necessary features are rather special, which explains why QCs are rare in nature. These are best seen by considering the particle pair interaction potentials and pair correlation functions in Fourier space, where one observes that there are two characteristic peaks at wave numbers $k_1$ and $k_2$, with the ratio $k_1/k_2$ taking certain special values which are geometric in origin \cite{lifshitz_PRL_97, lifshitz_07, 2D_2, barkan2014controlled, 3D_1, andy_PRL_13, 2D_3, 2D_1, walters2018structural, sbl18, dan}; e.g.\ for two dimensional (2D) dodecagonal QCs, $k_1/k_2=2\cos(\pi/12)\approx 1.93$. These features also manifest in the dispersion relation $\omega(k)$, which characterises the growth or decay of density modulations with wave number $k$ in the liquid state.

For one-component systems in the liquid state with number density $\rho_0$ we may express a perturbation in the density profile $\delta\rho(\textbf{r},t)\equiv\rho(\textbf{r},t)-\rho_0$ as a Fourier sum of modes with wave vector $\textbf{k}$ of form $\sim \exp(i\textbf{k}\cdot\textbf{r}+\omega t)$. The equation for the time evolution of the density $\rho$ can be written $\partial_t\delta\rho=\mathcal{L}\delta\rho+\mathcal{O}(\delta\rho^2)$, where $\partial_t$ is a partial time derivative and $\mathcal{L}$ is a spatial operator \cite{archer2012solidification, dispersion_relation, archer2016generation}. Linearising, we see that $\omega$ is the eigenvalue of $\mathcal{L}$ acting on the eigenfunction $\exp(i\textbf{k}\cdot\textbf{r})$, so if $\omega < 0$ for all $k=|\textbf{k}|$, then all modes decay and the uniform state is linearly stable. However, if $\omega > 0$ for some $k$, then those modes grow over time. How $\omega(k)$ is related to the form of the soft pair potentials is well established \cite{evans1979nature, archer2004dynamical, archer2012solidification, dispersion_relation, archer2016generation}.

To obtain QCs, the dispersion relation should exhibit maxima at $k_1$ and $k_2$, with roughly equal peak values that are as close to zero as possible. This is equivalent to requiring that the static structure factor $S(k)$ \cite{hansen} should exhibit two prominent peaks at $k_1$ and $k_2$, since for a bulk fluid $\omega(k)\propto-k^2/S(k)$ \cite{archer2004dynamical, andy_PRL_13, 2D_1, walters2018structural, archer2006dynamical}. Additionally, $\omega(k)$ must be sufficiently negative at the reciprocal lattice vectors of $k_1$ and $k_2$ that are involved in stabilising competing periodic crystal structures (e.g.\ with wavenumbers $\sqrt{3}k_1$, $\sqrt{3}k_2$, etc), so that these are suppressed \cite{dan}.

In addition to characterising how a uniform bulk liquid evolves after being perturbed, the dispersion relation is important because it gives crucial understanding of what wave number density modulations are favourable and which are likely be present in any incipient nonuniform crystalline or QC states \cite{dan}. In systems that are near to or beyond freezing, the characteristic modes that form the crystal or QC either grow or decay slowly. In one-component QC forming systems, this occurs in the vicinity of the point in the phase diagram where the system is marginally unstable at both $k_1$ and $k_2$  \cite{lifshitz_PRL_97, lifshitz_07, 2D_2, barkan2014controlled, 3D_1, andy_PRL_13, 2D_3, 2D_1, walters2018structural, sbl18, dan}.  

On the face of it, QCs should occur more widely in two component systems, since these intrinsically have at least two lengthscales, originating from the different particle sizes. Indeed, the vast majority of QCs discovered so far are metallic alloys with at least two components, e.g.\ Al-Mn or Ni-Cr \cite{discover, ishimasa1985new}. For mixtures where the particles have a well defined (hard) core, requiring certain geometrical motifs as minimal energy structures in local particle arrangements can be a fruitful way to find QCs \cite{widom1987quasicrystal, leung1989dodecagonal, talapin2009quasicrystalline, salgado2015non}. However, there is not an established `recipe' for finding them, at least in soft matter. The three-step strategy we follow and advocate here, which works for the colloidal mixture model considered below and which we expect to be more generally applicable, is as follows: (i) Obtain the liquid state partial static structure factors. (ii) Select the parameters or state point such that they exhibit two peaks at wave numbers $k_1$ and $k_2$ with the specific ratio corresponding to the desired QC, whilst also checking that there is no peak at $k=0$, as this is a signature of demixing, which can overtake the desired QC formation. (iii) Tune the parameters or state point so that the maxima in $\omega(k)$ are similar in height and as close to zero as possible. We successfully apply this strategy for finding QCs in a binary mixture modelled using a simple classical density functional theory (DFT) with direct correlations functions obtained from the hypernetted chain (HNC) Ornstein-Zernike integral equation theory \cite{hansen}. This is only qualitatively correct for the system to which it is applied (see below), but the simplicity makes it an ideal test system on which to develop our approach.

Our strategy has similarities to the approach used previously to find QCs in one-component systems, but for binary mixtures there are several additional complexities to overcome. Mixtures of big ($b$) and small ($s$) particles generically have at least three (not two) lengthscales present. These are the characteristic ranges of the $b$-$b$ and $s$-$s$ interactions and also the $b$-$s$ cross interaction. In the beautiful work \cite{barkan_thesis}, results for several model binary mixtures of soft particles are presented, treated using a phase field crystal type theory \cite{emmerich2012phase}. This predicts the mixtures to form a variety of QCs, with the $b$-$b$ and $s$-$s$ interaction potentials providing the two lengthscales needed for the QC formation. However, treating these mixtures with a more accurate DFT, which retains the logarithmic ideal gas free energy instead of approximating via a Taylor expansion \cite{emmerich2012phase, archer2019deriving}, we find that these systems actually just phase separate and do not form QCs (results not displayed). In the mixture considered here, the two lengthscales required for QC formation originate in the $b$-$b$ and $b$-$s$ interaction lengthscales. The $s$-$s$ lengthscale seems to be irrelevant.

\begin{figure}[t!]
\includegraphics[width=1.\columnwidth]{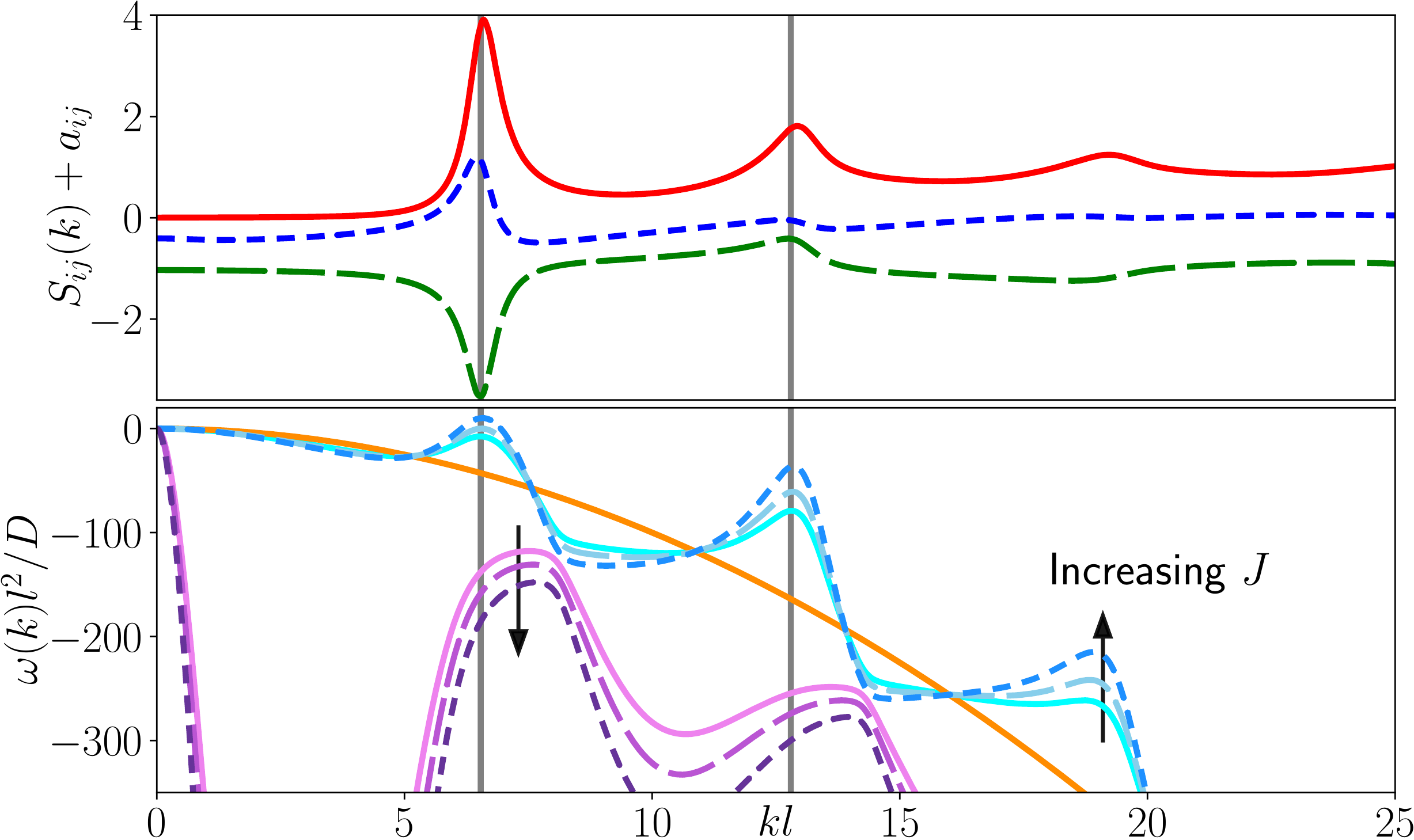}
\caption{Top: the three partial structure factors $S_{ij}(k)+a_{ij}$, where $a_{ij}$ is a constant shift, for clarity. The solid line is $S_{bb}(k)+0$, the long-dashed line is $S_{bs}(k)-1$, and the short-dashed line is $S_{ss}(k)-1$. The densities are $\rho_{0,b}l^2=1$ and $\rho_{0,s}l^2=2$ and the pair potential parameters are $\Gamma=42$, $m_s=0.025$ and $J=1$. The vertical lines correspond to the two characteristic lengthscales required for dodecagonal QCs. Bottom: the two branches of the dispersion relation, $\omega_{+}(k)$ and $\omega_{-}(k)$, for $J=1$, 1.25 and 1.5. The diffusion coefficients $D_b=D_s=D$ (see the SI). The solid orange line is the ideal gas result.}\label{structure}
\end{figure}

To determine the growth or decay rate of density perturbations in a uniform binary fluid mixture (i.e.\ the dispersion relation), where the bulk densities of the two species are $\rho_{0,b}$ and $\rho_{0,s}$, one must consider the time evolution of $\delta\boldsymbol{\rho}=(\delta\rho_b,\delta\rho_s)=(\rho_b-\rho_{0,b},\rho_s-\rho_{0,s})$. Fourier transforming the coupled dynamical equations for the two density profiles we obtain $\partial_t\hat{\boldsymbol{\rho}}(\textbf{k},t)=\textbf{L}\hat{\boldsymbol{\rho}}(\textbf{k},t)+\mathcal{O}(\hat{\boldsymbol{\rho}}^2)$, where $\hat{\boldsymbol{\rho}}$ is the Fourier transform of $\delta\boldsymbol{\rho}$ and $\textbf{L}$ is a $2\times2$ matrix \cite{dispersion_relation}. In the Appendix we give an explicit expression for $\textbf{L}$, which depends on $\hat{c}_{ij}(k)$, the Fourier transforms of the fluid pair direct correlation functions $c_{ij}(r)$, where $i,j=b,s$, for the case when the particles have Brownian equations of motion, i.e.\ where we can use dynamical DFT to describe the dynamics \cite{archer2004dynamical, Marconi:TarazonaJCP1999, mt00, ar04, Archer05}. For binary mixtures, the dispersion relation has two distinct branches, $\omega_{+}(k)$ and $\omega_{-}(k)$. The three partial structure factors $S_{ij}(k)$ are closely linked to $\hat{c}_{ij}(k)$ \cite{hansen} (see the Appendix) and therefore $\omega_+$ and $\omega_-$ depend crucially on the form of $S_{ij}(k)$. In Fig.~\ref{structure} we display examples of $S_{ij}(k)$ and also $\omega_{+}(k)$ and $\omega_{-}(k)$ for the model defined below. Note that the peak locations in $S_{ij}(k)$ are where the peaks in $\omega_+$ occur, i.e.\ these are the wave numbers of the slowest decaying density modes (or growing, if $\omega_+(k)>0$). The lower panel of Fig.~\ref{structure} shows how $\omega_+$ and $\omega_-$ vary as the inverse-temperature-like parameter $J$ (defined below) is varied. We compare these with the ideal gas case, where $\omega(k)\propto-k^2$, which highlights how the particle interactions are responsible for the shape of $\omega_+$ and $\omega_-$. Since $\omega_+\geq\omega_{-}$, the most important branch is $\omega_{+}$.

\begin{figure*}[!t]
  \includegraphics[width=0.77\linewidth]{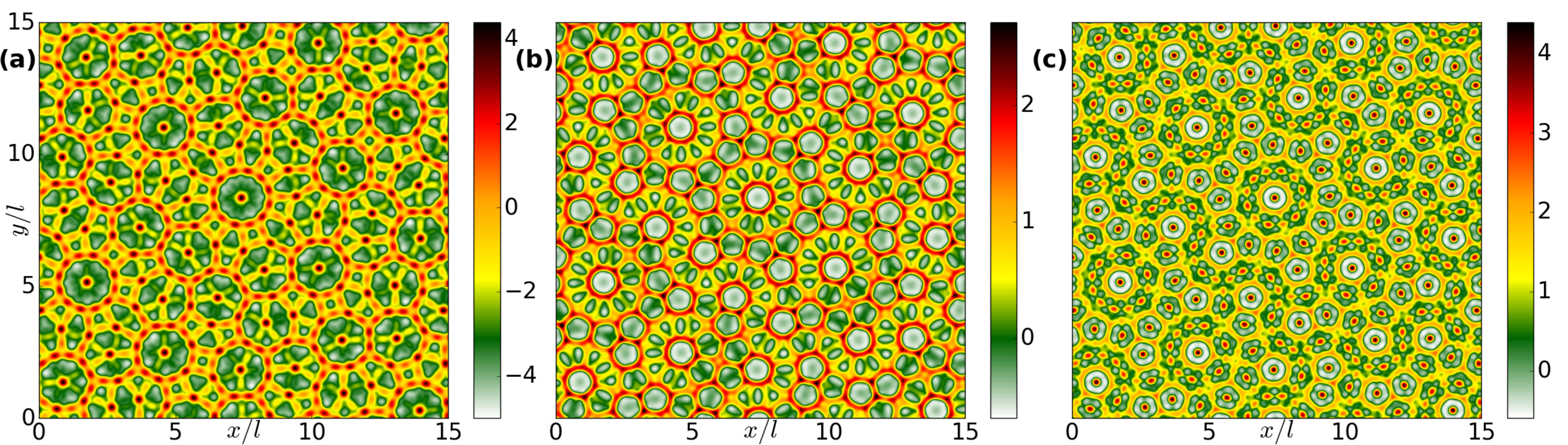}
  \includegraphics[width=0.22\linewidth]{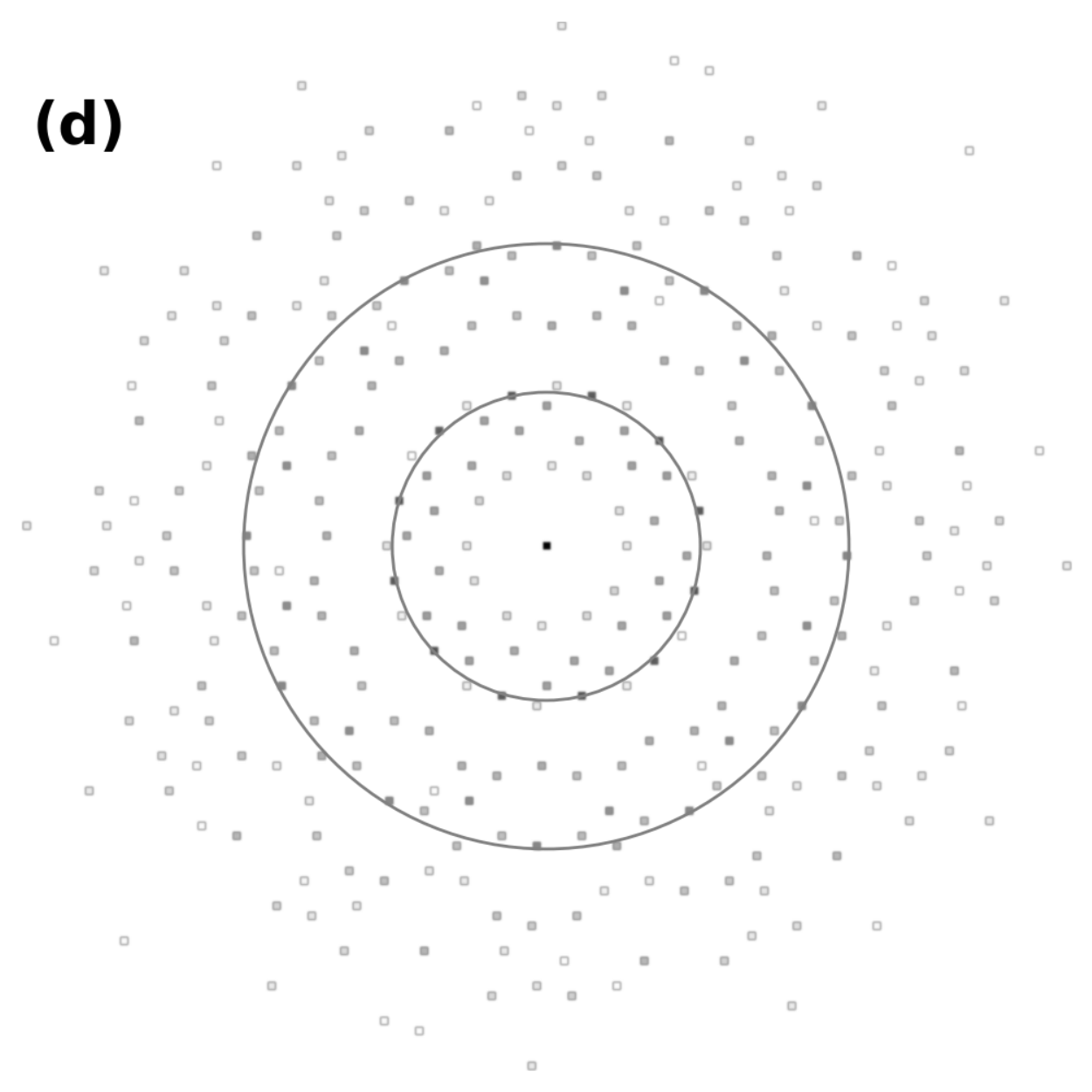}
  \caption{The logarithm of the two densities and the total density: in (a) we display $\ln(\rho_bl^2)$, in (b) $\ln(\rho_sl^2)$ and in (c) $\ln(\rho_bl^2+\rho_sl^2)$. In (d) is the Fourier transform of the $b$-particle density. The highest peaks in (a) and (c) correspond to the positions of the $b$-particles, while the $s$-particles in (b) are much more delocalised and almost fluid-like. The state point is the same as in Fig.~\ref{structure}, with $J=1$. In (d), the inner circle has radius $k_1$ and the outer has radius $k_2=1.93 k_1$.}\label{full_case}
\end{figure*}

The system we consider is a 2D binary mixture of charged colloids adsorbed on a flat oil-water interface \cite{Bresme2007, Law2011, Law2011b, walter}. The interactions between particles can be modelled by the pair potentials
\begin{equation}\label{eq:1}
\beta \phi_{ij}(r)= \Gamma m_i m_j\frac{l^3}{r^3},
\end{equation}
where $r$ is the distance between particles and $\beta=1/k_BT$, where $k_B$ is Boltzmann's constant and $T$ is the temperature. $l=1/\sqrt{\rho_{0,b}}$ is the typical distance between $b$-particles, $\Gamma$ is the dimensionless interaction strength between $b$-particles and $m_i$ is the dipole moment ratio of species $i=s,b$ relative to that of species $b$ (i.e.\ $m_b=1$ and $m_s<1$). This system exhibits a rich variety of 2D crystal structures \cite{Stirner2005, Assoud2007, Fornleitner2009, Chremos2009, Law2011, Law2011b, walter}. 

\begin{figure}[t!]
  \includegraphics[width=.98\columnwidth]{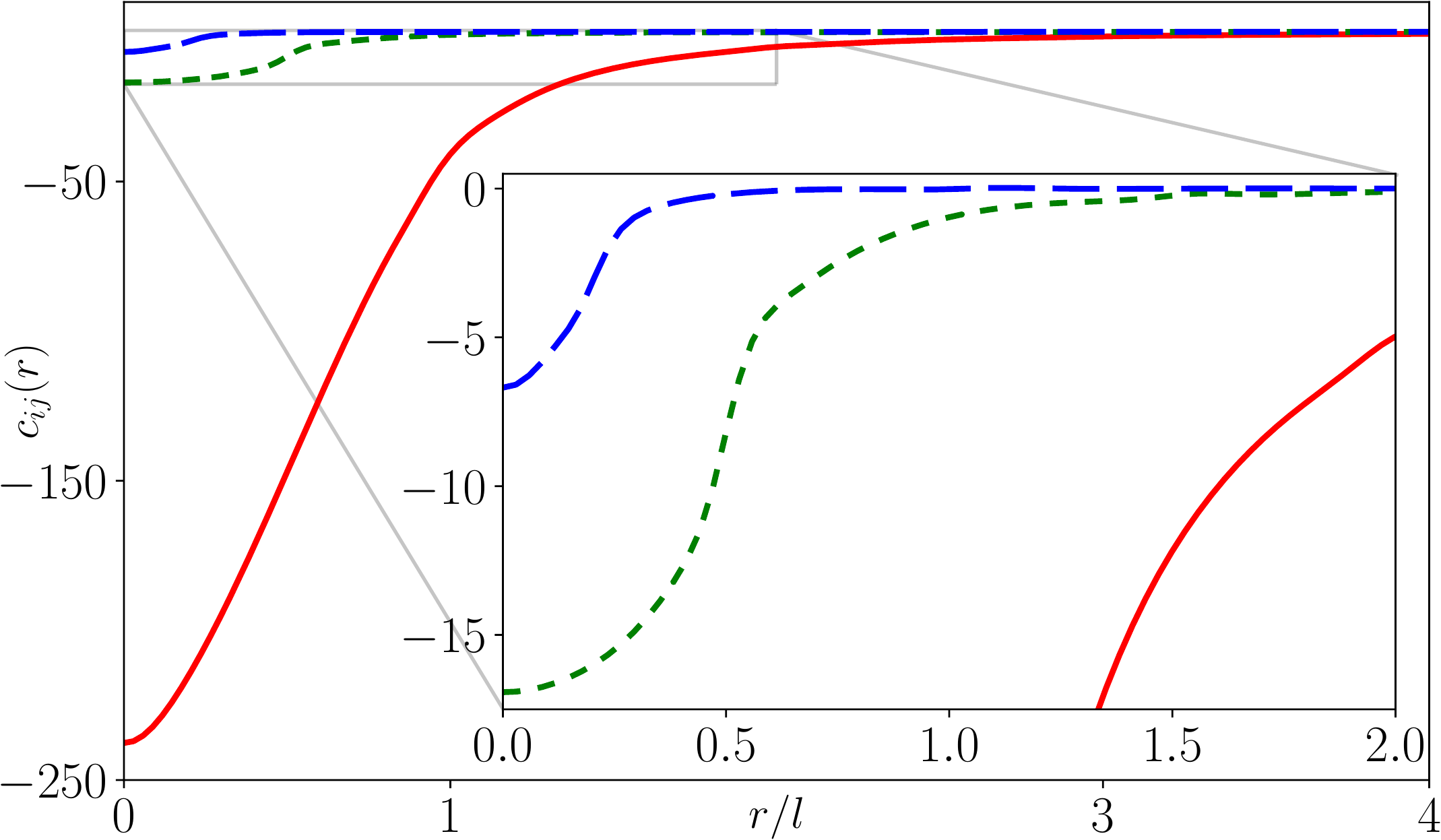}
  \caption{Pair direct correlation functions $c_{bb}(r)$ (red solid line), $c_{bs}(r)$ (short-dashed green line) and $c_{ss}(r)$ (dashed blue line) obtained from HNC theory, for the same state point as the results in Figs.~\ref{structure} and \ref{full_case}. The inset displays a magnification for small $r$. These are inputs for our DFT calculations.}\label{cij}
\end{figure}

We describe the system using DFT with the Ramakrishnan-Yussouff (RY) approximation \cite{ramakrishnan1979first}. The RY DFT was used in Ref.~\cite{walter} to obtain density profiles for system \eqref{eq:1} at state points where periodic crystals occur, with the $c_{ij}(r)$ that are inputs to the DFT obtained from an accurate but computationally intensive theory. Here instead the $c_{ij}(r)$ are obtained from HNC theory \cite{hansen}, which is less accurate but simpler and so much faster \cite{walter}. The RY DFT and HNC theory are described in the Appendix. The speed of this is important because tuning the pair potential parameters following the three steps of the recipe above for finding QCs requires numerous calculations. So, although the theory we use is at best qualitatively accurate \cite{walter}, it is used because of its speed and the fact that our aim is to test the efficacy of the strategy for finding QCs, not the accuracy of the theory. If more accuracy were required, one could use the DFT in Refs.~\cite{likos_lowen_06, likos_lowen_08}.

Density profiles for a QC state are displayed in Fig.~\ref{full_case}, with parameters $\rho_{0,b}l^2=1$, $\rho_{0,s}l^2=2$, $\Gamma=42$ and $m_s=0.025$, identified by following the three-step strategy given above. The partial structure factors and dispersion relations displayed in Fig.~\ref{structure} also correspond to this system. The branch $\omega_+(k)$ has a peak at $k=k_1$ with $\omega_+(k_1)$ approaching zero from below, but the second peak at $k_2>k_1$ is not as high. The pair potential parameters were tuned so that $k_1/k_2\approx1.93$, in order to observe dodecagonal QCs. The aim was of course to have both peaks at the same height and as close to zero as possible, but the physical constraints stemming from the form of the potentials in Eq.~\eqref{eq:1} prevents this. In particular, $\phi_{bs}(r)$ cannot be varied independently of $\phi_{bb}(r)$ and $\phi_{ss}(r)$. Of course, the HNC theory also fails before $\omega_+(k)\to0^-$.

To obtain the QC density profiles after calculating the pair direct correlation functions, in the DFT we initially replace $c_{ij}(r)\to J c_{ij}(r)$, where $J$ is a constant scaling parameter. Recall that $c_{ij}(r)\sim-\beta\phi_{ij}(r)$ for large $r$ \cite{hansen}, so increasing $J$ is much like decreasing the temperature. This makes freezing easier and the uniform density state to be linearly unstable, so using an initial guess for the two density profiles consisting of the desired average density value plus a small amplitude random field is sufficient to observe the QC formation (or periodic crystals, at other state points). See Ref.~\cite{walter} for further details on this approach involving $J$ for calculating solid state density profiles. To obtain the results in Fig.~\ref{full_case}, we initially calculate for $J=1.5$ and then take the resulting profiles as our initial guess at the physical value $J=1$. The density profile for the $b$-particles exhibits sharp peaks, with the QC structure clearly visible. The Fourier transform in Fig.~\ref{full_case}(d) shows the characteristic 12-fold symmetry. In contrast, the $s$-particles are much more delocalised and fluid-like, acting as a `stabiliser' for the structure. For these particles, the density peaks represent preferred locations where particles might be found some of the time, as there are more such locations than particles. Similar behaviour was observed at other state-points, where periodic crystals are the equilibria \cite{walter}.

\begin{figure}[t!]
\includegraphics[width=1.\columnwidth]{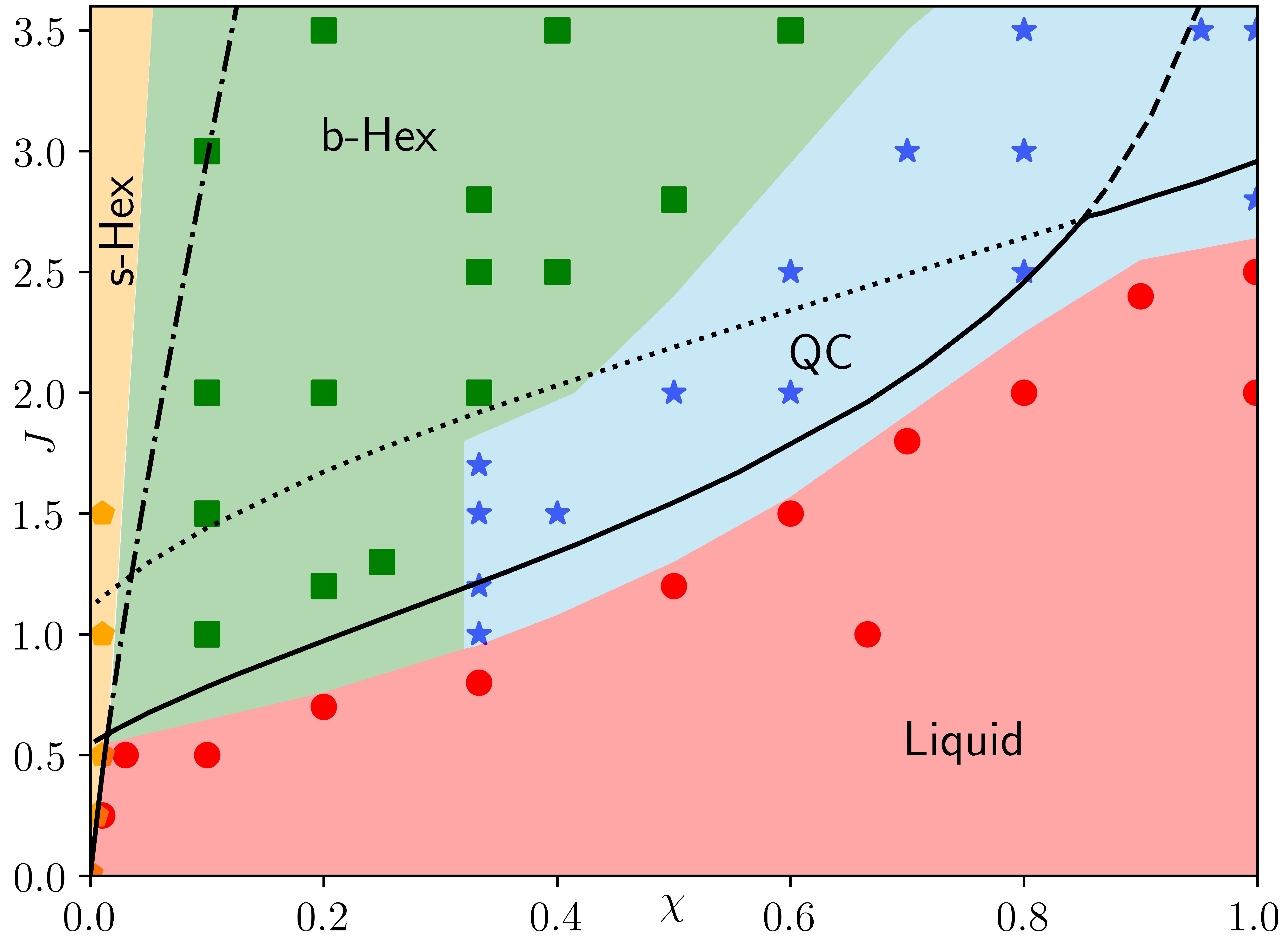}
\caption{Phase diagram in the $J$ versus concentration $\chi$ plane. Four phases are observed: liquid (circles), small lattice spacing crystal ($s$-Hex, pentagons), large lattice spacing crystal ($b$-Hex, squares) and QCs (stars). The boundaries between each are guides to the eye. The liquid state is unstable above the solid line stability threshold, where $\omega(k)$ has a maximum for $k\neq0$ with $\omega(k)=0$. The dotted, dashed and dot-dash lines are prolongations, along which $\omega(k)=0$ at one maximum, whilst already being positive at another.}\label{phase_diagram}
\end{figure}

\begin{figure}[t!]
  \includegraphics[width=1\columnwidth]{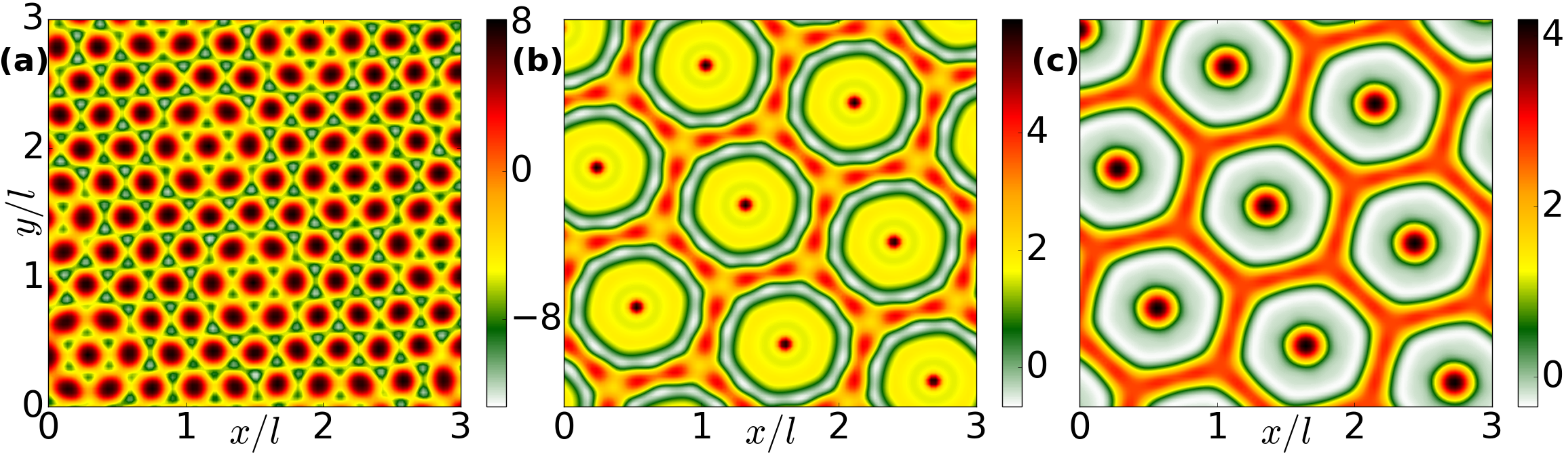}
  \caption{Logarithm of the total density for the periodic phases in the phase diagram, Fig.~\ref{phase_diagram}. (a) $s$-Hex at $(\chi,J)=(0.01,1)$. (b) $b$-Hex at $(\chi,J)=(0.1,1.5)$, with localised $s$-particles. (c) $b$-Hex at $(\chi,J)=(0.2,1.2)$, with $s$-particles free to move on a honeycomb like structure around the frozen $b$-particles.}\label{hexagons}
\end{figure}

The $c_{ij}(r)$ obtained from the HNC theory are displayed in Fig.~\ref{cij}. Inspecting these, one can roughly identify a typical lengthscale (effective diameter) as the range $r$ beyond which $c_{ij}(r)/c_{ij}(0)$ becomes small. The effective diameter obtained from $c_{bb}(r)$ is $\approx l$ and from $c_{bs}(r)$ is $\approx0.5l$. The ratio of these is $\approx2$, but one needs to go to Fourier space (Fig.~\ref{structure}) to see much more precisely the ratio $k_1/k_2\approx 1.93$, characteristic of QC formation. Note too that the effective diameter from $c_{ss}(r)$ is $\approx 0.25 l$. This corresponds to a wave vector $k_{s}l \approx 2\pi/0.25 \approx 25$. Fig.~\ref{structure} shows that neither $\omega(k)$ nor $S_{ij}(k)$ have significant features near $k_s$, indicating that this lengthscale is irrelevant to the QC formation.

A deeper understanding of the observed QC formation can be obtained by considering the phase diagram in the concentration $\chi\equiv\rho_{0,b}/(\rho_{0,b}+\rho_{0,s})$ versus $J$ plane, calculated using only the (scaled with $J$) pair direct correlation functions displayed in Fig.~\ref{cij} from the state point $(\chi,J)=(1/3,1)$ as input to the RY DFT. The result is displayed in Fig.~\ref{phase_diagram}. We should emphasise that because the phase diagram is calculated by rescaling the $c_{ij}(r)$ from the state point $(\chi,J)=(1/3,1)$ to all other state points, in a strict sense, this is the only physically relevant state point in Fig.~\ref{phase_diagram}. However, by exploring this theoretical model phase diagram, we obtain important insight into the observed QC formation that we would not obtain otherwise. At small $J$ the uniform density liquid state is found; recall that decreasing $J$ is like increasing $T$. At higher $J$ the system freezes to form one of three different solid phases. For small $\chi$, i.e.\ where the $s$-particles dominate, the system forms a hexagonal crystal with lattice spacing $\approx2\pi/k_{s}$, which we refer to as $s$-Hex; see Fig.~\ref{phase_diagram}. A portion of a typical example is displayed in Fig.~\ref{hexagons}(a). The defects originate from the the random initial conditions. Increasing $\chi$, the system forms a hexagonal crystal with much larger lattice spacing $\approx 2\pi/k_1$, determined by the range of $c_{bb}(r)$ ($b$-Hex in Fig.~\ref{phase_diagram}). These crystal structures are discussed in detail in \cite{Law2011}. See also the DFT results in \cite{walter}. In these, the $s$-particles can either be fluid-like, leading to a honeycomb like density distribution surrounding the peaks of the $b$-particles -- see Fig.~\ref{hexagons}(c). Alternatively, they can be more localised, so that the $b$-particle density peaks are surrounded by density peaks from the $s$-particles -- see Fig.~\ref{hexagons}(b). Moving to even higher $\chi$, we find QCs. An example is displayed in Fig.~\ref{full_case}. Note that in this model [obtained by rescaling the $c_{ij}(r)$ from the state point $(\chi,J)=(1/3,1)$] the QCs extends right up to $\chi=1$ (where $\rho_{0,s}=0$). We believe this is because the influence of the $s$-particles is still present in the rescaled $c_{bb}(r)$ that is calculated at $(\chi,J)=(1/3,1)$, although it could also be because $-c_{bb}(r)$ is somewhat akin to the soft effective pair potential of the monodisperse system in Ref.~\cite{zu2017forming}, which forms QCs.

This simplified model enables us to easily calculate the linear stability threshold for the uniform liquid, i.e.\ the locus in the phase diagram where either $\omega_+(k_1)=0$ or $\omega_+(k_2)=0$, or $\omega_+(k_{s})=0$. These are the lines in Fig.~\ref{phase_diagram}. Those satisfying the first two of these conditions meet at $(\chi,J)=(0.85,2.715)$, where the system is marginally unstable at both $k_1$ and $k_2$ (right hand cusp on the solid line in Fig.~\ref{phase_diagram}). Recall that for monodisperse systems points of this type are intimately connected with QC formation \cite{lifshitz_PRL_97, lifshitz_07, 2D_2, barkan2014controlled, 3D_1, andy_PRL_13, 2D_3, 2D_1, walters2018structural, sbl18, dan}. Thus, finding this point explains much of why we observe QCs in the present binary mixture. Importantly, notice that this point exists in the theoretically-constructed $(\chi,J)$ plane, rather than in the physical parameter space of the original system. So, although in the physical parameter space the two peaks at $k_1$ and $k_2$ in $\omega(k)$ are not at the same height and nor is the second peak in $S_{ij}(k)$ at $k_2$ as prominent as the first at $k_1$, nonetheless there is still the influence of the interaction between density modes with wavenumber $k_1$ and $k_2$ to stabilise the QC state. Much insight on such two-mode interactions is in the pattern formation literature related to Faraday waves \cite{lifshitz_PRL_97, sbl18, Edwards1994, Gollub1995, Besson1996, zhang1997pattern, kudrolli1998superlattice, silber2000two,  arbell2002pattern, porter2004pattern, porter2004resonant, ding2006enhanced, skeldon2007pattern, rucklidge2009design, rucklidge2012three, skeldon2015can} and though binary mixtures have the added complication of consisting of two coupled fields, much of this insight still applies.

In Fig.~\ref{phase_diagram} the boundaries between regions of the different phases are only guides for the eye. For $J=2$ we have determined the states at coexistence between the $b$-Hex and QC phases and found the width of the coexistence region to be $\Delta \chi \approx 0.04$ (not displayed). Since this is small, it justifies our approximate approach for identifying the locations of the phase boundaries (see the Appendix).

To summarise, we have proposed a `recipe' for finding QCs in soft matter mixtures. The key quantities for inspection are the partial static structure factors and the dispersion relation. In the model system studied here the two lengthscales required for QC formation arise from the $b$-$b$ and $b$-$s$ particle interactions. In principle, these could instead arise from the $b$-$b$ and $s$-$s$ interactions, but from our studies of soft-particle models (not shown), phase separation occurs much more readily than QC formation in this case. Using the cross-interaction $b$-$s$ lengthscale as one of the key QC lengthscales helps to avoid this.

This work is supported by the Swiss National Science Foundation under the grant number P2FRP2$\_$181453 (A.S.), and by the EPSRC under grants EP/L025078/1 (W.R.C.S. and D.M.A.B.) and EP/P015689/1 (A.J.A.). We thank Ron Lifshitz, Daniel Ratliff, Alastair Rucklidge, and Priya Subramanian for fruitful discussions at various points over the course of this work. We also thank Ron Lifshitz for sending us Ref.~\cite{barkan_thesis}.

\section*{Appendix}


\subsection*{The Ornstein-Zernike equation and liquid state structure}

The Ornstein-Zernike (OZ) equations for the total correlation functions $h_{ij}(r)$ of a binary fluid mixture are:
\begin{equation}
h_{ij}(r)=c_{ij}(r)+\sum_{p=b,s}\rho_{0,p}\int d\textbf{r}' c_{ip}(|\textbf{r}-\textbf{r}'|)h_{pj}(\textbf{r}'),
\label{eq:OZ}
\end{equation}
where $c_{ij}(r)$ are the pair direct correlation functions and $\rho_{0,i}$ for $i=b,s$ are the bulk fluid densities of the two species \cite{hansen}. The radial distribution functions are related to the total correlation functions via $g_{ij}(r)=1+h_{ij}(r)$. These coupled equations must be solved in conjunction with the following (exact) closure relations
\begin{equation}
c_{ij}(r)=-\beta\phi_{ij}(r)+h_{ij}(r)-\ln(1+h_{ij}(r))+B_{ij}(r),
\label{eq:closure}
\end{equation}
where $B_{ij}(r)$ are the so-called bridge-functions, $\phi_{ij}(r)$ are the pair potentials and $\beta=1/k_BT$ \cite{hansen}. The hypernetted chain (HNC) approximation consists of setting $B_{ij}(r)=0$ for all $r$. Due to the convolutions in \eqref{eq:OZ}, on Fourier transforming we obtain the following set of algebraic equations
\begin{equation}
\hat{h}_{ij}(k)=\hat{c}_{ij}(k)+\sum_{p=b,s}\rho_{0,p}\hat{c}_{ip}(k)\hat{h}_{pj}(k),
\label{eq:OZ_FT}
\end{equation}
where $\hat{h}_{ij}(k)$ and $\hat{c}_{ij}(k)$ are the Fourier transforms of $h_{ij}(r)$ and $c_{ij}(r)$, respectively. The partial static structure factors are related to these as follows \cite{hansen, dispersion_relation}:
\begin{equation}
\begin{split}
S_{bb}(k)&=1+\rho_{0,b}\hat{h}_{bb}(k),\\
S_{ss}(k)&=1+\rho_{0,s}\hat{h}_{ss}(k),\\
S_{bs}(k)&=\sqrt{\rho_{0,b}\rho_{0,s}}\hat{h}_{bs}(k).
\end{split}
\end{equation}
From \eqref{eq:OZ_FT} we obtain
\begin{equation}
\hat{h}_{ij}(k)=\frac{N_{ij}(k)}{D(k)},
\end{equation}
with the numerators given by
\begin{equation}
\begin{split}
N_{bb}(k)&=\hat{c}_{bb}(k)+\rho_{0,s}\left[\hat{c}_{bs}^2(k)-\hat{c}_{bb}(k)\hat{c}_{ss}(k)\right],\\
N_{ss}(k)&=\hat{c}_{ss}(k)+\rho_{0,b}\left[\hat{c}_{bs}^2(k)-\hat{c}_{bb}(k)\hat{c}_{ss}(k)\right],\\
N_{bs}(k)&=\hat{c}_{bs}(k).
\end{split}
\end{equation}
and the common denominator
\begin{equation}\label{D_of_k}
D(k)\equiv \left[1-\rho_{0,b}\hat{c}_{bb}(k)\right]\left[1-\rho_{0,s}\hat{c}_{ss}(k)\right]-\rho_{0,b}\rho_{0,s}\hat{c}_{bs}^2(k).
\end{equation}
For the stable liquid, $D(k)>0$ for all $k$. However, if this is not the case, then the liquid state is unstable. Thus, we can determine the stability threshold for the uniform liquid from solving for the locus in the phase diagram where a solution to the equation $D(k)=0$ appears.

\subsection*{Density functional theory for binary mixtures}

The density profiles $\rho_i(\mathbf{r})$ are obtained using classical density functional theory (DFT). The grand potential of the system is \cite{hansen, evans1979nature}
\begin{equation}
\Omega[\rho_b,\rho_s]=\mathcal{F}[\rho_b, \rho_s]+\sum_{i=b,s}\int d\textbf{r}\left(V_{i}^{\textrm{ext}}(\textbf{r})-\mu_{i}\right)\rho_{i}(\textbf{r}),
\end{equation}
where $\mathcal{F}$ is the intrinsic Helmholtz free energy functional, $V_{i}^{\textrm{ext}}(\mathbf{r})$ is the one-body external potential acting on species $i$ (here we set $V_{i}^{\textrm{ext}}(\textbf{r})\equiv 0$ for $i=b,s$, in order to study bulk phases) and $\mu_{i}$ are the chemical potentials. The intrinsic Helmholtz free energy can be split into two terms
\begin{equation}\label{helmholtz}
\mathcal{F}[\rho_b,\rho_s]=\mathcal{F}^{\textrm{id}}[\rho_b,\rho_s]+\mathcal{F}^{\textrm{ex}}[\rho_b,\rho_s],
\end{equation}
where the first term is the ideal gas contribution,
\begin{equation}
\mathcal{F}^{\textrm{id}}[\rho_b,\rho_s]=k_B T\sum_{i=b,s}\int d\textbf{r} \rho_{i}(\textbf{r})\left[\ln(\Lambda_{i}^d\rho_{i}(\textbf{r})-1\right],
\end{equation}
where $\Lambda_{i}$ is the (irrelevant) thermal de Broglie wavelength and $d$ is the dimensionality of the system. The second term in Eq.~\eqref{helmholtz} is the excess Helmholtz free energy, arising from the interactions between the particles. Following Ramakrishnan and Yussouff \cite{ramakrishnan1979first}, the approximation we use here is to expand this functional around the homogeneous fluid state in a functional Taylor expansion and truncate at second order, giving
\begin{eqnarray}
\mathcal{F}^{\textrm{ex}}[\rho_b,\rho_s]=\mathcal{F}^{\textrm{ex}}[\rho_{0,b},\rho_{0,s}]
+\sum_{i=b,s}\int d\textbf{r}\mu^{\textrm{ex}}_{i}\delta\rho_{i}(\textbf{r})\nonumber\\
-\frac{1}{2\beta}\sum_{\substack{i=b,s\\ j=b,s}}\int d\textbf{r}\delta\rho_{i}(\textbf{r})c_{ij}(\mid \textbf{r}-\textbf{r}'\mid)\delta\rho_{j}(\textbf{r}'),
\end{eqnarray}
where $\delta\rho_{i}(\textbf{r})=\rho_{i}(\textbf{r})-\rho_{0,i}$ and $\mu^{\textrm{ex}}_{i}=\mu_{i}-k_B T\ln\left(\rho_{0,i}\Lambda_{i}^d\right)$ are the excess chemical potentials. We further approximate the pair direct correlation functions $c_{ij}(r)$ via those obtained from the HNC theory. The equilibrium density profiles are those which minimise the grand potential $\Omega$ and which therefore satisfy the following pair of coupled Euler-Lagrange equations
\begin{equation}
\frac{\delta\Omega[\rho_b,\rho_s]}{\delta\rho_{i}}=0,
\end{equation}
for $i=b,s$.

\subsection*{Dynamics: the growth or decay of small amplitude density perturbations}

When the equations of motion of the particles can be approximated by stochastic Brownian equations of motion, then dynamical density functional theory (DDFT) shows that the non-equilibrium density distributions for the two species of particles $\rho_i(\mathbf{r},t)$ is described by  \cite{archer2004dynamical, Marconi:TarazonaJCP1999, mt00, ar04}:
\begin{equation}
\frac{\partial \rho_i}{\partial t} =
\nabla\cdot\left(\gamma_i\rho_i\nabla
\frac{\delta \Omega[\rho_s,\rho_b]}{\delta\rho_i}\right),
\label{eq:DDFT}
\end{equation}
where the mobility coefficient $\gamma_i=\beta D_i$ and where $D_i$ is the diffusion coefficient of species $i$. Note that if instead the particles evolve according to Newton's equations of motion, then the equations for the time evolution of the density profiles are more complicated, but in dense systems one can argue that Eq.~\eqref{eq:DDFT} still governs the long time (on diffusive timescales) behaviour \cite{Archer05}. If we consider the growth or decay of small amplitude density perturbations around the bulk value of the form $\delta\rho_{i}(\textbf{r},t)=\rho_{i}(\textbf{r},t)-\rho_{0,i}$, then we can expand Eqs.~\eqref{eq:DDFT} to obtain \cite{evans1979nature, archer2004dynamical, archer2012solidification, dispersion_relation, archer2016generation}:
\begin{eqnarray}
\frac{\partial \delta\rho_i(\textbf{r},t)}{\partial t} =
D_i\nabla^2\delta\rho_i(\textbf{r},t)\hspace{4.5cm}\nonumber\\
-D_i\rho_{0,i}\sum_{j=b,s}\nabla^2\int d\textbf{r}'\delta\rho_{j}(\textbf{r}',t)c_{ij}(\mid \textbf{r}-\textbf{r}'\mid)\nonumber\\
+O(\delta \rho_i^2).\hspace{5.2cm}
\label{eq:DDFT_linear}
\end{eqnarray}
Linearising this equation and then Fourier transforming, we obtain
\begin{equation}
\frac{\partial \hat{\rho}_i(\mathbf{k},t)}{\partial t} =
-k^2D_i\hat{\rho}_i(\mathbf{k},t)+k^2D_i\rho_{0,i}\sum_{j=b,s}\hat{\rho}_{j}(\mathbf{k},t)c_{ij}(k),
\label{eq:DDFT_linear_FT}
\end{equation}
where $\hat{\rho}_i(\mathbf{k},t)$ is the Fourier transform of $\delta\rho_i(\mathbf{r},t)$ and $k=|\mathbf{k}|$. Assuming $\hat{\rho}_i(\mathbf{k},t)\propto\exp(\omega(k)t)$, then Eq.~\eqref{eq:DDFT_linear_FT} becomes \cite{dispersion_relation}:
\begin{equation}
\textbf{1}\omega(k)\hat{\boldsymbol{\rho}}=\textbf{L}\hat{\boldsymbol{\rho}},
\label{eq:matrix_eq}
\end{equation}
where $\hat{\boldsymbol{\rho}}=(\hat{\rho}_b,\hat{\rho}_s)$ and the matrix $\textbf{L}=\textbf{M}\textbf{E}$, where the two matrices $\textbf{M}$ and $\textbf{E}$ are defined as
\begin{equation}\label{eq:M_matrix}
\textbf{M}=-k^2\begin{pmatrix}
D_b\rho_{0,b} & 0\\
0 & D_s\rho_{0,s}
\end{pmatrix}
\end{equation}
and
\begin{equation}
\textbf{E}=\begin{pmatrix}
\left[\frac{1}{\rho_{0,b}}-\hat{c}_{bb}(k)\right] & -\hat{c}_{bs}(k)\\
-\hat{c}_{sb}(k) & \left[\frac{1}{\rho_{0,s}}-\hat{c}_{ss}(k)\right]
\end{pmatrix}.
\end{equation}
Solving Eq.~\eqref{eq:matrix_eq} for the dispersion relation $\omega(k)$, one obtains two branches of solutions, $\omega_{\pm}(k)$. These are given by
\begin{equation}\label{dispersion_relation_eq}
\omega_{\pm}(k)=\frac{1}{2}\textrm{Tr}(\textbf{M}\textbf{E})\pm \sqrt{\frac{1}{4}\textrm{Tr}(\textbf{M}\textbf{E})^2-\textrm{det}(\textbf{M}\textbf{E})}.
\end{equation}
Further details of this derivation can be found in Ref.~\cite{dispersion_relation}. Note that the equation $\textrm{det}(\textbf{E})=0$ is entirely equivalent to solving $D(k)=0$, from Eq.~\eqref{D_of_k}.

It is worth recalling that the values of the diffusion coefficients $D_b$ and $D_s$ do not ever determine which structure is the thermodynamic equilibrium state, i.e.\ the minimum of the free energy. Therefore, the values of $D_b$ and $D_s$ are not involved in determining the phase diagram in Fig.~4 of the main text. Nor do the values of $D_b$ and $D_s$ determine the locations of the linear stability threshold lines in the phase diagram, i.e.\ the lines in Fig.~4 where either $\omega_+(k_1)=0$ or $\omega_+(k_2)=0$ or $\omega_+(k_{s})=0$. This is because these lines come from solving the equation $\textrm{det}(\textbf{E})=0$, whilst the values of the diffusion coefficients only enter the mobility matrix $\textbf{M}$ in Eq.~\eqref{eq:M_matrix}. That said, the precise value of the ratio $D_b/D_s$ does influence the dispersion relation curves, but does not affect where the peaks occur (i.e.\ does not change $k_1$ or $k_2$). Thus, the value of the ratio $D_b/D_s$ is only relevant to the non-equilibrium dynamics of the system. However, since here we are solely ultimately interested in the equilibrium phase behaviour of the system, which does not depend on $D_b/D_s$, we therefore set this ratio equal to 1, i.e.\ we set $D_b=D_s=D$.

\subsection*{Note on the width of the coexistence region between the QC and the $b$-Hex phase}

In the main text we comment briefly on the fact that in the phase diagram in Fig.~4 the coexistence region between the QC and the $b$-Hex phase is fairly small. It is worth expanding on those comments here. That the coexistence region is narrow is important, because it implies that in large portions of the phase diagram (as displayed in Fig~4), the QC is the thermodynamic equilibrium. In the main text we give the width of the coexistence region $\Delta\chi\approx0.04$ for $J=2$. For lower values of $J$ the coexistence region becomes a little broader (e.g.\ at $J=1.5$ the width of the coexistence region $\Delta\chi\approx0.06$) and for higher $J$ it is narrower. Other model systems where the coexistence gap between the QC and hexagonal phases is very narrow include the systems described in Refs.~\cite{andy_PRL_13, dan}, so based on our experience with those systems, the narrowness in the present system is perhaps not too surprising.

Another observation on this issue worth noting is the following: If one initiates the system in the QC state and then decreases $\chi$ in small steps, following the QC branch of solutions, one eventually falls off that branch onto the $b$-Hex phase branch of solutions. For example, for $J=1.5$ this occurs at $\chi\approx0.3$. Some authors would refer to this as the ``spinodal'' point for the QC phase. In other words, for $J=1.5$ and $\chi<0.3$ the QC state is no longer a stable solution to the model equations. In a similar way, if one initiates the system in the $b$-Hex state and then increases $\chi$ in small steps, following the $b$-Hex branch of solutions, one eventually falls off that branch onto a state that is a periodic approximant for the QC state. For $J=1.5$ this $b$-Hex spinodal point occurs at $\chi\approx0.37$. In other words, for $J=1.5$ and $\chi>0.37$ the $b$-Hex state is no longer a stable solution to the model equations. This fact that the system falls from $b$-Hex branch of solutions onto a branch related to the QC state is a very strong indicator that the QC is the thermodynamic equilibrium state. Moreover, the distance in the phase diagram between these two spinodal points $0.37-0.30=0.07$, is an upper bound for the coexistence region width $\Delta\chi$.


\begin{thebibliography}{57}%
\makeatletter
\providecommand \@ifxundefined [1]{%
 \@ifx{#1\undefined}
}%
\providecommand \@ifnum [1]{%
 \ifnum #1\expandafter \@firstoftwo
 \else \expandafter \@secondoftwo
 \fi
}%
\providecommand \@ifx [1]{%
 \ifx #1\expandafter \@firstoftwo
 \else \expandafter \@secondoftwo
 \fi
}%
\providecommand \natexlab [1]{#1}%
\providecommand \enquote  [1]{``#1''}%
\providecommand \bibnamefont  [1]{#1}%
\providecommand \bibfnamefont [1]{#1}%
\providecommand \citenamefont [1]{#1}%
\providecommand \href@noop [0]{\@secondoftwo}%
\providecommand \href [0]{\begingroup \@sanitize@url \@href}%
\providecommand \@href[1]{\@@startlink{#1}\@@href}%
\providecommand \@@href[1]{\endgroup#1\@@endlink}%
\providecommand \@sanitize@url [0]{\catcode `\\12\catcode `\$12\catcode
  `\&12\catcode `\#12\catcode `\^12\catcode `\_12\catcode `\%12\relax}%
\providecommand \@@startlink[1]{}%
\providecommand \@@endlink[0]{}%
\providecommand \url  [0]{\begingroup\@sanitize@url \@url }%
\providecommand \@url [1]{\endgroup\@href {#1}{\urlprefix }}%
\providecommand \urlprefix  [0]{URL }%
\providecommand \Eprint [0]{\href }%
\providecommand \doibase [0]{http://dx.doi.org/}%
\providecommand \selectlanguage [0]{\@gobble}%
\providecommand \bibinfo  [0]{\@secondoftwo}%
\providecommand \bibfield  [0]{\@secondoftwo}%
\providecommand \translation [1]{[#1]}%
\providecommand \BibitemOpen [0]{}%
\providecommand \bibitemStop [0]{}%
\providecommand \bibitemNoStop [0]{.\EOS\space}%
\providecommand \EOS [0]{\spacefactor3000\relax}%
\providecommand \BibitemShut  [1]{\csname bibitem#1\endcsname}%
\let\auto@bib@innerbib\@empty
\bibitem [{\citenamefont {Lifshitz}\ and\ \citenamefont
  {Petrich}(1997)}]{lifshitz_PRL_97}%
  \BibitemOpen
  \bibfield  {author} {\bibinfo {author} {\bibfnamefont {R.}~\bibnamefont
  {Lifshitz}}\ and\ \bibinfo {author} {\bibfnamefont {D.~M.}\ \bibnamefont
  {Petrich}},\ }\bibfield  {title} {\enquote {\bibinfo {title} {Theoretical
  model for {F}araday waves with multiple-frequency forcing},}\ }\href@noop {}
  {\bibfield  {journal} {\bibinfo  {journal} {Phys. Rev. Lett.}\ }\textbf
  {\bibinfo {volume} {79}},\ \bibinfo {pages} {1261} (\bibinfo {year}
  {1997})}\BibitemShut {NoStop}%
\bibitem [{\citenamefont {Lifshitz}\ and\ \citenamefont
  {Diamant}(2007)}]{lifshitz_07}%
  \BibitemOpen
  \bibfield  {author} {\bibinfo {author} {\bibfnamefont {R.}~\bibnamefont
  {Lifshitz}}\ and\ \bibinfo {author} {\bibfnamefont {H.}~\bibnamefont
  {Diamant}},\ }\bibfield  {title} {\enquote {\bibinfo {title} {Soft
  quasicrystals--why are they stable?}}\ }\href@noop {} {\bibfield  {journal}
  {\bibinfo  {journal} {Philos. Mag.}\ }\textbf {\bibinfo {volume} {87}},\
  \bibinfo {pages} {3021} (\bibinfo {year} {2007})}\BibitemShut {NoStop}%
\bibitem [{\citenamefont {Barkan}\ \emph {et~al.}(2011)\citenamefont {Barkan},
  \citenamefont {Diamant},\ and\ \citenamefont {Lifshitz}}]{2D_2}%
  \BibitemOpen
  \bibfield  {author} {\bibinfo {author} {\bibfnamefont {K.}~\bibnamefont
  {Barkan}}, \bibinfo {author} {\bibfnamefont {H.}~\bibnamefont {Diamant}}, \
  and\ \bibinfo {author} {\bibfnamefont {R.}~\bibnamefont {Lifshitz}},\
  }\bibfield  {title} {\enquote {\bibinfo {title} {Stability of quasicrystals
  composed of soft isotropic particles},}\ }\href@noop {} {\bibfield  {journal}
  {\bibinfo  {journal} {Phys. Rev. B}\ }\textbf {\bibinfo {volume} {83}},\
  \bibinfo {pages} {172201} (\bibinfo {year} {2011})}\BibitemShut {NoStop}%
\bibitem [{\citenamefont {Barkan}\ \emph {et~al.}(2014)\citenamefont {Barkan},
  \citenamefont {Engel},\ and\ \citenamefont
  {Lifshitz}}]{barkan2014controlled}%
  \BibitemOpen
  \bibfield  {author} {\bibinfo {author} {\bibfnamefont {K.}~\bibnamefont
  {Barkan}}, \bibinfo {author} {\bibfnamefont {M.}~\bibnamefont {Engel}}, \
  and\ \bibinfo {author} {\bibfnamefont {R.}~\bibnamefont {Lifshitz}},\
  }\bibfield  {title} {\enquote {\bibinfo {title} {Controlled self-assembly of
  periodic and aperiodic cluster crystals},}\ }\href@noop {} {\bibfield
  {journal} {\bibinfo  {journal} {Phys. Rev. Lett.}\ }\textbf {\bibinfo
  {volume} {113}},\ \bibinfo {pages} {098304} (\bibinfo {year}
  {2014})}\BibitemShut {NoStop}%
\bibitem [{\citenamefont {Subramanian}\ \emph {et~al.}(2016)\citenamefont
  {Subramanian}, \citenamefont {Archer}, \citenamefont {Knobloch},\ and\
  \citenamefont {Rucklidge}}]{3D_1}%
  \BibitemOpen
  \bibfield  {author} {\bibinfo {author} {\bibfnamefont {P.}~\bibnamefont
  {Subramanian}}, \bibinfo {author} {\bibfnamefont {A.~J.}\ \bibnamefont
  {Archer}}, \bibinfo {author} {\bibfnamefont {E.}~\bibnamefont {Knobloch}}, \
  and\ \bibinfo {author} {\bibfnamefont {A.~M.}\ \bibnamefont {Rucklidge}},\
  }\bibfield  {title} {\enquote {\bibinfo {title} {Three-dimensional
  icosahedral phase field quasicrystal},}\ }\href@noop {} {\bibfield  {journal}
  {\bibinfo  {journal} {Phys. Rev. Lett.}\ }\textbf {\bibinfo {volume} {117}},\
  \bibinfo {pages} {075501} (\bibinfo {year} {2016})}\BibitemShut {NoStop}%
\bibitem [{\citenamefont {Archer}\ \emph {et~al.}(2013)\citenamefont {Archer},
  \citenamefont {Rucklidge},\ and\ \citenamefont {Knobloch}}]{andy_PRL_13}%
  \BibitemOpen
  \bibfield  {author} {\bibinfo {author} {\bibfnamefont {A.~J.}\ \bibnamefont
  {Archer}}, \bibinfo {author} {\bibfnamefont {A.~M.}\ \bibnamefont
  {Rucklidge}}, \ and\ \bibinfo {author} {\bibfnamefont {E.}~\bibnamefont
  {Knobloch}},\ }\bibfield  {title} {\enquote {\bibinfo {title}
  {Quasicrystalline order and a crystal-liquid state in a soft-core fluid},}\
  }\href@noop {} {\bibfield  {journal} {\bibinfo  {journal} {Phys. Rev. Lett.}\
  }\textbf {\bibinfo {volume} {111}},\ \bibinfo {pages} {165501} (\bibinfo
  {year} {2013})}\BibitemShut {NoStop}%
\bibitem [{\citenamefont {Dotera}\ \emph {et~al.}(2014)\citenamefont {Dotera},
  \citenamefont {Oshiro},\ and\ \citenamefont {Ziherl}}]{2D_3}%
  \BibitemOpen
  \bibfield  {author} {\bibinfo {author} {\bibfnamefont {T.}~\bibnamefont
  {Dotera}}, \bibinfo {author} {\bibfnamefont {T.}~\bibnamefont {Oshiro}}, \
  and\ \bibinfo {author} {\bibfnamefont {P.}~\bibnamefont {Ziherl}},\
  }\bibfield  {title} {\enquote {\bibinfo {title} {Mosaic two-lengthscale
  quasicrystals},}\ }\href@noop {} {\bibfield  {journal} {\bibinfo  {journal}
  {Nature}\ }\textbf {\bibinfo {volume} {506}},\ \bibinfo {pages} {208}
  (\bibinfo {year} {2014})}\BibitemShut {NoStop}%
\bibitem [{\citenamefont {Archer}\ \emph {et~al.}(2015)\citenamefont {Archer},
  \citenamefont {Rucklidge},\ and\ \citenamefont {Knobloch}}]{2D_1}%
  \BibitemOpen
  \bibfield  {author} {\bibinfo {author} {\bibfnamefont {A.~J.}\ \bibnamefont
  {Archer}}, \bibinfo {author} {\bibfnamefont {A.~M.}\ \bibnamefont
  {Rucklidge}}, \ and\ \bibinfo {author} {\bibfnamefont {E.}~\bibnamefont
  {Knobloch}},\ }\bibfield  {title} {\enquote {\bibinfo {title} {Soft-core
  particles freezing to form a quasicrystal and a crystal-liquid phase},}\
  }\href@noop {} {\bibfield  {journal} {\bibinfo  {journal} {Phys. Rev. E}\
  }\textbf {\bibinfo {volume} {92}},\ \bibinfo {pages} {012324} (\bibinfo
  {year} {2015})}\BibitemShut {NoStop}%
\bibitem [{\citenamefont {Walters}\ \emph {et~al.}(2018)\citenamefont
  {Walters}, \citenamefont {Subramanian}, \citenamefont {Archer},\ and\
  \citenamefont {Evans}}]{walters2018structural}%
  \BibitemOpen
  \bibfield  {author} {\bibinfo {author} {\bibfnamefont {M.~C.}\ \bibnamefont
  {Walters}}, \bibinfo {author} {\bibfnamefont {P.}~\bibnamefont
  {Subramanian}}, \bibinfo {author} {\bibfnamefont {A.~J.}\ \bibnamefont
  {Archer}}, \ and\ \bibinfo {author} {\bibfnamefont {R.}~\bibnamefont
  {Evans}},\ }\bibfield  {title} {\enquote {\bibinfo {title} {Structural
  crossover in a model fluid exhibiting two length scales: repercussions for
  quasicrystal formation},}\ }\href@noop {} {\bibfield  {journal} {\bibinfo
  {journal} {Phys. Rev. E}\ }\textbf {\bibinfo {volume} {98}},\ \bibinfo
  {pages} {012606} (\bibinfo {year} {2018})}\BibitemShut {NoStop}%
\bibitem [{\citenamefont {Savitz}\ \emph {et~al.}(2018)\citenamefont {Savitz},
  \citenamefont {Babadi},\ and\ \citenamefont {Lifshitz}}]{sbl18}%
  \BibitemOpen
  \bibfield  {author} {\bibinfo {author} {\bibfnamefont {S.}~\bibnamefont
  {Savitz}}, \bibinfo {author} {\bibfnamefont {M.}~\bibnamefont {Babadi}}, \
  and\ \bibinfo {author} {\bibfnamefont {R.}~\bibnamefont {Lifshitz}},\
  }\bibfield  {title} {\enquote {\bibinfo {title} {Multiple-scale structures:
  from {F}araday waves to soft-matter quasicrystals},}\ }\href@noop {}
  {\bibfield  {journal} {\bibinfo  {journal} {IUCrJ}\ }\textbf {\bibinfo
  {volume} {5}},\ \bibinfo {pages} {247} (\bibinfo {year} {2018})}\BibitemShut
  {NoStop}%
\bibitem [{\citenamefont {Ratliff}\ \emph {et~al.}(2019)\citenamefont
  {Ratliff}, \citenamefont {Archer}, \citenamefont {Subramanian},\ and\
  \citenamefont {Rucklidge}}]{dan}%
  \BibitemOpen
  \bibfield  {author} {\bibinfo {author} {\bibfnamefont {D.~J.}\ \bibnamefont
  {Ratliff}}, \bibinfo {author} {\bibfnamefont {A.~J.}\ \bibnamefont {Archer}},
  \bibinfo {author} {\bibfnamefont {P.}~\bibnamefont {Subramanian}}, \ and\
  \bibinfo {author} {\bibfnamefont {A.~M.}\ \bibnamefont {Rucklidge}},\
  }\bibfield  {title} {\enquote {\bibinfo {title} {Which wave numbers determine
  the thermodynamic stability of soft matter quasicrystals?}}\ }\href@noop {}
  {\bibfield  {journal} {\bibinfo  {journal} {Phys. Rev. Lett.}\ }\textbf
  {\bibinfo {volume} {123}},\ \bibinfo {pages} {148004} (\bibinfo {year}
  {2019})}\BibitemShut {NoStop}%
\bibitem [{\citenamefont {Archer}\ \emph {et~al.}(2012)\citenamefont {Archer},
  \citenamefont {Robbins}, \citenamefont {Thiele},\ and\ \citenamefont
  {Knobloch}}]{archer2012solidification}%
  \BibitemOpen
  \bibfield  {author} {\bibinfo {author} {\bibfnamefont {A.~J.}\ \bibnamefont
  {Archer}}, \bibinfo {author} {\bibfnamefont {M.~J.}\ \bibnamefont {Robbins}},
  \bibinfo {author} {\bibfnamefont {U.}~\bibnamefont {Thiele}}, \ and\ \bibinfo
  {author} {\bibfnamefont {E.}~\bibnamefont {Knobloch}},\ }\bibfield  {title}
  {\enquote {\bibinfo {title} {Solidification fronts in supercooled liquids:
  How rapid fronts can lead to disordered glassy solids},}\ }\href@noop {}
  {\bibfield  {journal} {\bibinfo  {journal} {Phys. Rev. E}\ }\textbf {\bibinfo
  {volume} {86}},\ \bibinfo {pages} {031603} (\bibinfo {year}
  {2012})}\BibitemShut {NoStop}%
\bibitem [{\citenamefont {Archer}\ \emph {et~al.}(2014)\citenamefont {Archer},
  \citenamefont {Walters}, \citenamefont {Thiele},\ and\ \citenamefont
  {Knobloch}}]{dispersion_relation}%
  \BibitemOpen
  \bibfield  {author} {\bibinfo {author} {\bibfnamefont {A.~J.}\ \bibnamefont
  {Archer}}, \bibinfo {author} {\bibfnamefont {M.~C.}\ \bibnamefont {Walters}},
  \bibinfo {author} {\bibfnamefont {U.}~\bibnamefont {Thiele}}, \ and\ \bibinfo
  {author} {\bibfnamefont {E.}~\bibnamefont {Knobloch}},\ }\bibfield  {title}
  {\enquote {\bibinfo {title} {Solidification in soft-core fluids: Disordered
  solids from fast solidification fronts},}\ }\href@noop {} {\bibfield
  {journal} {\bibinfo  {journal} {Phys. Rev. E}\ }\textbf {\bibinfo {volume}
  {90}},\ \bibinfo {pages} {042404} (\bibinfo {year} {2014})}\BibitemShut
  {NoStop}%
\bibitem [{\citenamefont {Archer}\ \emph {et~al.}(2016)\citenamefont {Archer},
  \citenamefont {Walters}, \citenamefont {Thiele},\ and\ \citenamefont
  {Knobloch}}]{archer2016generation}%
  \BibitemOpen
  \bibfield  {author} {\bibinfo {author} {\bibfnamefont {A.~J.}\ \bibnamefont
  {Archer}}, \bibinfo {author} {\bibfnamefont {M.~C.}\ \bibnamefont {Walters}},
  \bibinfo {author} {\bibfnamefont {U.}~\bibnamefont {Thiele}}, \ and\ \bibinfo
  {author} {\bibfnamefont {E.}~\bibnamefont {Knobloch}},\ }\bibfield  {title}
  {\enquote {\bibinfo {title} {Generation of defects and disorder from deeply
  quenching a liquid to form a solid},}\ }in\ \href@noop {} {\emph {\bibinfo
  {booktitle} {Mathematical Challenges in a New Phase of Materials Science}}}\
  (\bibinfo  {publisher} {Springer},\ \bibinfo {year} {2016})\ pp.\ \bibinfo
  {pages} {1--26}\BibitemShut {NoStop}%
\bibitem [{\citenamefont {Evans}(1979)}]{evans1979nature}%
  \BibitemOpen
  \bibfield  {author} {\bibinfo {author} {\bibfnamefont {R.}~\bibnamefont
  {Evans}},\ }\bibfield  {title} {\enquote {\bibinfo {title} {The nature of the
  liquid-vapour interface and other topics in the statistical mechanics of
  non-uniform, classical fluids},}\ }\href@noop {} {\bibfield  {journal}
  {\bibinfo  {journal} {Adv. Phys.}\ }\textbf {\bibinfo {volume} {28}},\
  \bibinfo {pages} {143} (\bibinfo {year} {1979})}\BibitemShut {NoStop}%
\bibitem [{\citenamefont {Archer}\ and\ \citenamefont
  {Evans}(2004)}]{archer2004dynamical}%
  \BibitemOpen
  \bibfield  {author} {\bibinfo {author} {\bibfnamefont {A.~J.}\ \bibnamefont
  {Archer}}\ and\ \bibinfo {author} {\bibfnamefont {R.}~\bibnamefont {Evans}},\
  }\bibfield  {title} {\enquote {\bibinfo {title} {Dynamical density functional
  theory and its application to spinodal decomposition},}\ }\href@noop {}
  {\bibfield  {journal} {\bibinfo  {journal} {J. Chem. Phys.}\ }\textbf
  {\bibinfo {volume} {121}},\ \bibinfo {pages} {4246} (\bibinfo {year}
  {2004})}\BibitemShut {NoStop}%
\bibitem [{\citenamefont {Hansen}\ and\ \citenamefont
  {McDonald}(2013)}]{hansen}%
  \BibitemOpen
  \bibfield  {author} {\bibinfo {author} {\bibfnamefont {J.-P.}\ \bibnamefont
  {Hansen}}\ and\ \bibinfo {author} {\bibfnamefont {I.~R.}\ \bibnamefont
  {McDonald}},\ }\href@noop {} {\emph {\bibinfo {title} {Theory of simple
  liquids: with applications to soft matter}}}\ (\bibinfo  {publisher}
  {Academic Press},\ \bibinfo {year} {2013})\BibitemShut {NoStop}%
\bibitem [{\citenamefont {Archer}(2006)}]{archer2006dynamical}%
  \BibitemOpen
  \bibfield  {author} {\bibinfo {author} {\bibfnamefont {A.~J.}\ \bibnamefont
  {Archer}},\ }\bibfield  {title} {\enquote {\bibinfo {title} {Dynamical
  density functional theory for dense atomic liquids},}\ }\href@noop {}
  {\bibfield  {journal} {\bibinfo  {journal} {J. Phys.: Condens. Matter}\
  }\textbf {\bibinfo {volume} {18}},\ \bibinfo {pages} {5617} (\bibinfo {year}
  {2006})}\BibitemShut {NoStop}%
\bibitem [{\citenamefont {Shechtman}\ \emph {et~al.}(1984)\citenamefont
  {Shechtman}, \citenamefont {Blech}, \citenamefont {Gratias},\ and\
  \citenamefont {Cahn}}]{discover}%
  \BibitemOpen
  \bibfield  {author} {\bibinfo {author} {\bibfnamefont {D.}~\bibnamefont
  {Shechtman}}, \bibinfo {author} {\bibfnamefont {I.}~\bibnamefont {Blech}},
  \bibinfo {author} {\bibfnamefont {D.}~\bibnamefont {Gratias}}, \ and\
  \bibinfo {author} {\bibfnamefont {J.~W.}\ \bibnamefont {Cahn}},\ }\bibfield
  {title} {\enquote {\bibinfo {title} {Metallic phase with long-range
  orientational order and no translational symmetry},}\ }\href@noop {}
  {\bibfield  {journal} {\bibinfo  {journal} {Phys. Rev. Lett.}\ }\textbf
  {\bibinfo {volume} {53}},\ \bibinfo {pages} {1951} (\bibinfo {year}
  {1984})}\BibitemShut {NoStop}%
\bibitem [{\citenamefont {Ishimasa}\ \emph {et~al.}(1985)\citenamefont
  {Ishimasa}, \citenamefont {Nissen},\ and\ \citenamefont
  {Fukano}}]{ishimasa1985new}%
  \BibitemOpen
  \bibfield  {author} {\bibinfo {author} {\bibfnamefont {T.}~\bibnamefont
  {Ishimasa}}, \bibinfo {author} {\bibfnamefont {H.-U.}\ \bibnamefont
  {Nissen}}, \ and\ \bibinfo {author} {\bibfnamefont {Y.}~\bibnamefont
  {Fukano}},\ }\bibfield  {title} {\enquote {\bibinfo {title} {New ordered
  state between crystalline and amorphous in {N}i-{C}r particles},}\
  }\href@noop {} {\bibfield  {journal} {\bibinfo  {journal} {Phys. Rev. Lett.}\
  }\textbf {\bibinfo {volume} {55}},\ \bibinfo {pages} {511} (\bibinfo {year}
  {1985})}\BibitemShut {NoStop}%
\bibitem [{\citenamefont {Widom}\ \emph {et~al.}(1987)\citenamefont {Widom},
  \citenamefont {Strandburg},\ and\ \citenamefont
  {Swendsen}}]{widom1987quasicrystal}%
  \BibitemOpen
  \bibfield  {author} {\bibinfo {author} {\bibfnamefont {M.}~\bibnamefont
  {Widom}}, \bibinfo {author} {\bibfnamefont {K.~J.}\ \bibnamefont
  {Strandburg}}, \ and\ \bibinfo {author} {\bibfnamefont {R.~H.}\ \bibnamefont
  {Swendsen}},\ }\bibfield  {title} {\enquote {\bibinfo {title} {Quasicrystal
  equilibrium state},}\ }\href@noop {} {\bibfield  {journal} {\bibinfo
  {journal} {Phys. Rev. Lett.}\ }\textbf {\bibinfo {volume} {58}},\ \bibinfo
  {pages} {706} (\bibinfo {year} {1987})}\BibitemShut {NoStop}%
\bibitem [{\citenamefont {Leung}\ \emph {et~al.}(1989)\citenamefont {Leung},
  \citenamefont {Henley},\ and\ \citenamefont
  {Chester}}]{leung1989dodecagonal}%
  \BibitemOpen
  \bibfield  {author} {\bibinfo {author} {\bibfnamefont {P.~W.}\ \bibnamefont
  {Leung}}, \bibinfo {author} {\bibfnamefont {C.~L.}\ \bibnamefont {Henley}}, \
  and\ \bibinfo {author} {\bibfnamefont {G.~V.}\ \bibnamefont {Chester}},\
  }\bibfield  {title} {\enquote {\bibinfo {title} {Dodecagonal order in a
  two-dimensional {L}ennard-{J}ones system},}\ }\href@noop {} {\bibfield
  {journal} {\bibinfo  {journal} {Phys. Rev. B}\ }\textbf {\bibinfo {volume}
  {39}},\ \bibinfo {pages} {446} (\bibinfo {year} {1989})}\BibitemShut
  {NoStop}%
\bibitem [{\citenamefont {Talapin}\ \emph {et~al.}(2009)\citenamefont
  {Talapin}, \citenamefont {Shevchenko}, \citenamefont {Bodnarchuk},
  \citenamefont {Ye}, \citenamefont {Chen},\ and\ \citenamefont
  {Murray}}]{talapin2009quasicrystalline}%
  \BibitemOpen
  \bibfield  {author} {\bibinfo {author} {\bibfnamefont {D.~V.}\ \bibnamefont
  {Talapin}}, \bibinfo {author} {\bibfnamefont {E.~V.}\ \bibnamefont
  {Shevchenko}}, \bibinfo {author} {\bibfnamefont {M.~I.}\ \bibnamefont
  {Bodnarchuk}}, \bibinfo {author} {\bibfnamefont {X.}~\bibnamefont {Ye}},
  \bibinfo {author} {\bibfnamefont {J.}~\bibnamefont {Chen}}, \ and\ \bibinfo
  {author} {\bibfnamefont {C.~B.}\ \bibnamefont {Murray}},\ }\bibfield  {title}
  {\enquote {\bibinfo {title} {Quasicrystalline order in self-assembled binary
  nanoparticle superlattices},}\ }\href@noop {} {\bibfield  {journal} {\bibinfo
   {journal} {Nature}\ }\textbf {\bibinfo {volume} {461}},\ \bibinfo {pages}
  {964} (\bibinfo {year} {2009})}\BibitemShut {NoStop}%
\bibitem [{\citenamefont {Salgado-Blanco}\ and\ \citenamefont
  {Mendoza}(2015)}]{salgado2015non}%
  \BibitemOpen
  \bibfield  {author} {\bibinfo {author} {\bibfnamefont {D.}~\bibnamefont
  {Salgado-Blanco}}\ and\ \bibinfo {author} {\bibfnamefont {C.~I.}\
  \bibnamefont {Mendoza}},\ }\bibfield  {title} {\enquote {\bibinfo {title}
  {Non-additive simple potentials for pre-programmed self-assembly},}\
  }\href@noop {} {\bibfield  {journal} {\bibinfo  {journal} {Soft Matter}\
  }\textbf {\bibinfo {volume} {11}},\ \bibinfo {pages} {889} (\bibinfo {year}
  {2015})}\BibitemShut {NoStop}%
\bibitem [{\citenamefont {Barkan}(2015)}]{barkan_thesis}%
  \BibitemOpen
  \bibfield  {author} {\bibinfo {author} {\bibfnamefont {K.}~\bibnamefont
  {Barkan}},\ }\emph {\bibinfo {title} {Theory and Simulation of the Self
  Assembly of Soft Quasicrystals.}},\ \href@noop {} {Ph.D. thesis},\ \bibinfo
  {school} {Tel Aviv University, Israel} (\bibinfo {year} {2015})\BibitemShut
  {NoStop}%
\bibitem [{\citenamefont {Emmerich}\ \emph {et~al.}(2012)\citenamefont
  {Emmerich}, \citenamefont {L{\"o}wen}, \citenamefont {Wittkowski},
  \citenamefont {Gruhn}, \citenamefont {T{\'o}th}, \citenamefont {Tegze},\ and\
  \citenamefont {Gr{\'a}n{\'a}sy}}]{emmerich2012phase}%
  \BibitemOpen
  \bibfield  {author} {\bibinfo {author} {\bibfnamefont {H.}~\bibnamefont
  {Emmerich}}, \bibinfo {author} {\bibfnamefont {H.}~\bibnamefont {L{\"o}wen}},
  \bibinfo {author} {\bibfnamefont {R.}~\bibnamefont {Wittkowski}}, \bibinfo
  {author} {\bibfnamefont {T.}~\bibnamefont {Gruhn}}, \bibinfo {author}
  {\bibfnamefont {G.~I.}\ \bibnamefont {T{\'o}th}}, \bibinfo {author}
  {\bibfnamefont {G.}~\bibnamefont {Tegze}}, \ and\ \bibinfo {author}
  {\bibfnamefont {L.}~\bibnamefont {Gr{\'a}n{\'a}sy}},\ }\bibfield  {title}
  {\enquote {\bibinfo {title} {Phase-field-crystal models for condensed matter
  dynamics on atomic length and diffusive time scales: an overview},}\
  }\href@noop {} {\bibfield  {journal} {\bibinfo  {journal} {Adv. Physics}\
  }\textbf {\bibinfo {volume} {61}},\ \bibinfo {pages} {665} (\bibinfo {year}
  {2012})}\BibitemShut {NoStop}%
\bibitem [{\citenamefont {Archer}\ \emph {et~al.}(2019)\citenamefont {Archer},
  \citenamefont {Ratliff}, \citenamefont {Rucklidge},\ and\ \citenamefont
  {Subramanian}}]{archer2019deriving}%
  \BibitemOpen
  \bibfield  {author} {\bibinfo {author} {\bibfnamefont {A.~J.}\ \bibnamefont
  {Archer}}, \bibinfo {author} {\bibfnamefont {D.~J.}\ \bibnamefont {Ratliff}},
  \bibinfo {author} {\bibfnamefont {A.~M.}\ \bibnamefont {Rucklidge}}, \ and\
  \bibinfo {author} {\bibfnamefont {P.}~\bibnamefont {Subramanian}},\
  }\bibfield  {title} {\enquote {\bibinfo {title} {Deriving phase field crystal
  theory from dynamical density functional theory: consequences of the
  approximations},}\ }\href@noop {} {\bibfield  {journal} {\bibinfo  {journal}
  {Phys. Rev. E}\ }\textbf {\bibinfo {volume} {100}},\ \bibinfo {pages}
  {022140} (\bibinfo {year} {2019})}\BibitemShut {NoStop}%
\bibitem [{\citenamefont {Marconi}\ and\ \citenamefont
  {Tarazona}(1999)}]{Marconi:TarazonaJCP1999}%
  \BibitemOpen
  \bibfield  {author} {\bibinfo {author} {\bibfnamefont {U.~M.~B.}\
  \bibnamefont {Marconi}}\ and\ \bibinfo {author} {\bibfnamefont
  {P.}~\bibnamefont {Tarazona}},\ }\bibfield  {title} {\enquote {\bibinfo
  {title} {Dynamic density functional theory of fluids},}\ }\href@noop {}
  {\bibfield  {journal} {\bibinfo  {journal} {J. Chem. Phys.}\ }\textbf
  {\bibinfo {volume} {110}},\ \bibinfo {pages} {8032} (\bibinfo {year}
  {1999})}\BibitemShut {NoStop}%
\bibitem [{\citenamefont {Marconi}\ and\ \citenamefont
  {Tarazona}(2000)}]{mt00}%
  \BibitemOpen
  \bibfield  {author} {\bibinfo {author} {\bibfnamefont {U.~M.~B.}\
  \bibnamefont {Marconi}}\ and\ \bibinfo {author} {\bibfnamefont
  {P.}~\bibnamefont {Tarazona}},\ }\bibfield  {title} {\enquote {\bibinfo
  {title} {Dynamic density functional theory of fluids},}\ }\href@noop {}
  {\bibfield  {journal} {\bibinfo  {journal} {J. Phys. Condens. Matt.}\
  }\textbf {\bibinfo {volume} {12}},\ \bibinfo {pages} {A413} (\bibinfo {year}
  {2000})}\BibitemShut {NoStop}%
\bibitem [{\citenamefont {Archer}\ and\ \citenamefont {Rauscher}(2004)}]{ar04}%
  \BibitemOpen
  \bibfield  {author} {\bibinfo {author} {\bibfnamefont {A.~J.}\ \bibnamefont
  {Archer}}\ and\ \bibinfo {author} {\bibfnamefont {M.}~\bibnamefont
  {Rauscher}},\ }\bibfield  {title} {\enquote {\bibinfo {title} {Dynamical
  density functional theory for interacting brownian particles: stochastic or
  deterministic?}}\ }\href@noop {} {\bibfield  {journal} {\bibinfo  {journal}
  {J. Phys. A}\ }\textbf {\bibinfo {volume} {37}},\ \bibinfo {pages} {9325}
  (\bibinfo {year} {2004})}\BibitemShut {NoStop}%
\bibitem [{\citenamefont {Archer}(2005)}]{Archer05}%
  \BibitemOpen
  \bibfield  {author} {\bibinfo {author} {\bibfnamefont {A.~J.}\ \bibnamefont
  {Archer}},\ }\bibfield  {title} {\enquote {\bibinfo {title} {Dynamical
  density functional theory: binary phase-separating colloidal fluid in a
  cavity},}\ }\href@noop {} {\bibfield  {journal} {\bibinfo  {journal} {J.
  Phys.: Cond. Mat.}\ }\textbf {\bibinfo {volume} {17}},\ \bibinfo {pages}
  {1405} (\bibinfo {year} {2005})}\BibitemShut {NoStop}%
\bibitem [{\citenamefont {Bresme}\ and\ \citenamefont
  {Oettel}(2007)}]{Bresme2007}%
  \BibitemOpen
  \bibfield  {author} {\bibinfo {author} {\bibfnamefont {F.}~\bibnamefont
  {Bresme}}\ and\ \bibinfo {author} {\bibfnamefont {M.}~\bibnamefont
  {Oettel}},\ }\bibfield  {title} {\enquote {\bibinfo {title} {Nanoparticles at
  fluid interfaces},}\ }\href@noop {} {\bibfield  {journal} {\bibinfo
  {journal} {J. Phys.: Condens. Matter}\ }\textbf {\bibinfo {volume} {19}},\
  \bibinfo {pages} {413101} (\bibinfo {year} {2007})}\BibitemShut {NoStop}%
\bibitem [{\citenamefont {Law}\ \emph {et~al.}(2011{\natexlab{a}})\citenamefont
  {Law}, \citenamefont {Buzza},\ and\ \citenamefont {Horozov}}]{Law2011}%
  \BibitemOpen
  \bibfield  {author} {\bibinfo {author} {\bibfnamefont {A.~D.}\ \bibnamefont
  {Law}}, \bibinfo {author} {\bibfnamefont {D.~M.~A.}\ \bibnamefont {Buzza}}, \
  and\ \bibinfo {author} {\bibfnamefont {T.~S.}\ \bibnamefont {Horozov}},\
  }\bibfield  {title} {\enquote {\bibinfo {title} {Two-dimensional colloidal
  alloys},}\ }\href@noop {} {\bibfield  {journal} {\bibinfo  {journal} {Phys.
  Rev. Lett.}\ }\textbf {\bibinfo {volume} {106}},\ \bibinfo {pages} {128302}
  (\bibinfo {year} {2011}{\natexlab{a}})}\BibitemShut {NoStop}%
\bibitem [{\citenamefont {Law}\ \emph {et~al.}(2011{\natexlab{b}})\citenamefont
  {Law}, \citenamefont {Horozov},\ and\ \citenamefont {Buzza}}]{Law2011b}%
  \BibitemOpen
  \bibfield  {author} {\bibinfo {author} {\bibfnamefont {A.~D.}\ \bibnamefont
  {Law}}, \bibinfo {author} {\bibfnamefont {T.~S.}\ \bibnamefont {Horozov}}, \
  and\ \bibinfo {author} {\bibfnamefont {D.~M.~A.}\ \bibnamefont {Buzza}},\
  }\bibfield  {title} {\enquote {\bibinfo {title} {The structure and melting
  transition of two-dimensional colloidal alloys},}\ }\href@noop {} {\bibfield
  {journal} {\bibinfo  {journal} {Soft Matter}\ }\textbf {\bibinfo {volume}
  {7}},\ \bibinfo {pages} {8923--8931} (\bibinfo {year}
  {2011}{\natexlab{b}})}\BibitemShut {NoStop}%
\bibitem [{\citenamefont {Somerville}\ \emph {et~al.}(2018)\citenamefont
  {Somerville}, \citenamefont {Stokes}, \citenamefont {Adawi}, \citenamefont
  {Horozov}, \citenamefont {Archer},\ and\ \citenamefont {Buzza}}]{walter}%
  \BibitemOpen
  \bibfield  {author} {\bibinfo {author} {\bibfnamefont {W.~R.~C.}\
  \bibnamefont {Somerville}}, \bibinfo {author} {\bibfnamefont {J.~L.}\
  \bibnamefont {Stokes}}, \bibinfo {author} {\bibfnamefont {A.~M.}\
  \bibnamefont {Adawi}}, \bibinfo {author} {\bibfnamefont {T.~S.}\ \bibnamefont
  {Horozov}}, \bibinfo {author} {\bibfnamefont {A.~J.}\ \bibnamefont {Archer}},
  \ and\ \bibinfo {author} {\bibfnamefont {D.~M.~A.}\ \bibnamefont {Buzza}},\
  }\bibfield  {title} {\enquote {\bibinfo {title} {Density functional theory
  for the crystallization of two-dimensional dipolar colloidal alloys},}\
  }\href@noop {} {\bibfield  {journal} {\bibinfo  {journal} {J. Phys.: Condens.
  Matter}\ }\textbf {\bibinfo {volume} {30}},\ \bibinfo {pages} {405102}
  (\bibinfo {year} {2018})}\BibitemShut {NoStop}%
\bibitem [{\citenamefont {Stirner}\ and\ \citenamefont
  {Sun}(2005)}]{Stirner2005}%
  \BibitemOpen
  \bibfield  {author} {\bibinfo {author} {\bibfnamefont {T.}~\bibnamefont
  {Stirner}}\ and\ \bibinfo {author} {\bibfnamefont {J.}~\bibnamefont {Sun}},\
  }\bibfield  {title} {\enquote {\bibinfo {title} {Molecular dynamics
  simulation of the structural configuration of binary colloidal monolayers},}\
  }\href@noop {} {\bibfield  {journal} {\bibinfo  {journal} {Langmuir}\
  }\textbf {\bibinfo {volume} {21}},\ \bibinfo {pages} {6636} (\bibinfo {year}
  {2005})}\BibitemShut {NoStop}%
\bibitem [{\citenamefont {Assoud}\ \emph {et~al.}(2007)\citenamefont {Assoud},
  \citenamefont {Messina},\ and\ \citenamefont {L{\"o}wen}}]{Assoud2007}%
  \BibitemOpen
  \bibfield  {author} {\bibinfo {author} {\bibfnamefont {L.}~\bibnamefont
  {Assoud}}, \bibinfo {author} {\bibfnamefont {R.}~\bibnamefont {Messina}}, \
  and\ \bibinfo {author} {\bibfnamefont {H.}~\bibnamefont {L{\"o}wen}},\
  }\bibfield  {title} {\enquote {\bibinfo {title} {Stable crystalline lattices
  in two-dimensional binary mixtures of dipolar particles},}\ }\href@noop {}
  {\bibfield  {journal} {\bibinfo  {journal} {EPL (Europhysics Letters)}\
  }\textbf {\bibinfo {volume} {80}},\ \bibinfo {pages} {48001} (\bibinfo {year}
  {2007})}\BibitemShut {NoStop}%
\bibitem [{\citenamefont {Fornleitner}\ \emph {et~al.}(2009)\citenamefont
  {Fornleitner}, \citenamefont {Lo~Verso}, \citenamefont {Kahl},\ and\
  \citenamefont {Likos}}]{Fornleitner2009}%
  \BibitemOpen
  \bibfield  {author} {\bibinfo {author} {\bibfnamefont {J.}~\bibnamefont
  {Fornleitner}}, \bibinfo {author} {\bibfnamefont {F.}~\bibnamefont
  {Lo~Verso}}, \bibinfo {author} {\bibfnamefont {G.}~\bibnamefont {Kahl}}, \
  and\ \bibinfo {author} {\bibfnamefont {C.~N.}\ \bibnamefont {Likos}},\
  }\bibfield  {title} {\enquote {\bibinfo {title} {Ordering in two-dimensional
  dipolar mixtures},}\ }\href@noop {} {\bibfield  {journal} {\bibinfo
  {journal} {Langmuir}\ }\textbf {\bibinfo {volume} {25}},\ \bibinfo {pages}
  {7836} (\bibinfo {year} {2009})}\BibitemShut {NoStop}%
\bibitem [{\citenamefont {Chremos}\ and\ \citenamefont
  {Likos}(2009)}]{Chremos2009}%
  \BibitemOpen
  \bibfield  {author} {\bibinfo {author} {\bibfnamefont {A.}~\bibnamefont
  {Chremos}}\ and\ \bibinfo {author} {\bibfnamefont {C.~N.}\ \bibnamefont
  {Likos}},\ }\bibfield  {title} {\enquote {\bibinfo {title} {Crystal
  structures of two-dimensional binary mixtures of dipolar colloids in tilted
  external magnetic fields},}\ }\href@noop {} {\bibfield  {journal} {\bibinfo
  {journal} {J. Phys. Chem. B}\ }\textbf {\bibinfo {volume} {113}},\ \bibinfo
  {pages} {12316} (\bibinfo {year} {2009})}\BibitemShut {NoStop}%
\bibitem [{\citenamefont {Ramakrishnan}\ and\ \citenamefont
  {Yussouff}(1979)}]{ramakrishnan1979first}%
  \BibitemOpen
  \bibfield  {author} {\bibinfo {author} {\bibfnamefont {T.~V.}\ \bibnamefont
  {Ramakrishnan}}\ and\ \bibinfo {author} {\bibfnamefont {M.}~\bibnamefont
  {Yussouff}},\ }\bibfield  {title} {\enquote {\bibinfo {title}
  {First-principles order-parameter theory of freezing},}\ }\href@noop {}
  {\bibfield  {journal} {\bibinfo  {journal} {Phys. Rev. B}\ }\textbf {\bibinfo
  {volume} {19}},\ \bibinfo {pages} {2775} (\bibinfo {year}
  {1979})}\BibitemShut {NoStop}%
\bibitem [{\citenamefont {van Teeffelen}\ \emph {et~al.}(2006)\citenamefont
  {van Teeffelen}, \citenamefont {Likos}, \citenamefont {Hoffmann},\ and\
  \citenamefont {L{\"o}wen}}]{likos_lowen_06}%
  \BibitemOpen
  \bibfield  {author} {\bibinfo {author} {\bibfnamefont {S.}~\bibnamefont {van
  Teeffelen}}, \bibinfo {author} {\bibfnamefont {C.~N.}\ \bibnamefont {Likos}},
  \bibinfo {author} {\bibfnamefont {N.}~\bibnamefont {Hoffmann}}, \ and\
  \bibinfo {author} {\bibfnamefont {H.}~\bibnamefont {L{\"o}wen}},\ }\bibfield
  {title} {\enquote {\bibinfo {title} {Density functional theory of freezing
  for soft interactions in two dimensions},}\ }\href@noop {} {\bibfield
  {journal} {\bibinfo  {journal} {EPL (Europhysics Letters)}\ }\textbf
  {\bibinfo {volume} {75}},\ \bibinfo {pages} {583} (\bibinfo {year}
  {2006})}\BibitemShut {NoStop}%
\bibitem [{\citenamefont {Van~Teeffelen}\ \emph {et~al.}(2008)\citenamefont
  {Van~Teeffelen}, \citenamefont {L{\"o}wen},\ and\ \citenamefont
  {Likos}}]{likos_lowen_08}%
  \BibitemOpen
  \bibfield  {author} {\bibinfo {author} {\bibfnamefont {S.}~\bibnamefont
  {Van~Teeffelen}}, \bibinfo {author} {\bibfnamefont {H.}~\bibnamefont
  {L{\"o}wen}}, \ and\ \bibinfo {author} {\bibfnamefont {C.~N.}\ \bibnamefont
  {Likos}},\ }\bibfield  {title} {\enquote {\bibinfo {title} {Crystallization
  of magnetic dipolar monolayers: a density functional approach},}\ }\href@noop
  {} {\bibfield  {journal} {\bibinfo  {journal} {J. Phys.: Condens. Matter}\
  }\textbf {\bibinfo {volume} {20}},\ \bibinfo {pages} {404217} (\bibinfo
  {year} {2008})}\BibitemShut {NoStop}%
\bibitem [{\citenamefont {Zu}\ \emph {et~al.}(2017)\citenamefont {Zu},
  \citenamefont {Tan},\ and\ \citenamefont {Xu}}]{zu2017forming}%
  \BibitemOpen
  \bibfield  {author} {\bibinfo {author} {\bibfnamefont {M.}~\bibnamefont
  {Zu}}, \bibinfo {author} {\bibfnamefont {P.}~\bibnamefont {Tan}}, \ and\
  \bibinfo {author} {\bibfnamefont {N.}~\bibnamefont {Xu}},\ }\bibfield
  {title} {\enquote {\bibinfo {title} {Forming quasicrystals by monodisperse
  soft core particles},}\ }\href@noop {} {\bibfield  {journal} {\bibinfo
  {journal} {Nature Comm.}\ }\textbf {\bibinfo {volume} {8}},\ \bibinfo {pages}
  {2089} (\bibinfo {year} {2017})}\BibitemShut {NoStop}%
\bibitem [{\citenamefont {Edwards}\ and\ \citenamefont
  {Fauve}(1994)}]{Edwards1994}%
  \BibitemOpen
  \bibfield  {author} {\bibinfo {author} {\bibfnamefont {W.~S.}\ \bibnamefont
  {Edwards}}\ and\ \bibinfo {author} {\bibfnamefont {S.}~\bibnamefont
  {Fauve}},\ }\bibfield  {title} {\enquote {\bibinfo {title} {Patterns and
  quasi-patterns in the {F}araday experiment},}\ }\href@noop {} {\bibfield
  {journal} {\bibinfo  {journal} {J. Fluid Mech.}\ }\textbf {\bibinfo {volume}
  {278}},\ \bibinfo {pages} {123} (\bibinfo {year} {1994})}\BibitemShut
  {NoStop}%
\bibitem [{\citenamefont {Gollub}(1995)}]{Gollub1995}%
  \BibitemOpen
  \bibfield  {author} {\bibinfo {author} {\bibfnamefont {J.~P.}\ \bibnamefont
  {Gollub}},\ }\bibfield  {title} {\enquote {\bibinfo {title} {Order and
  disorder in fluid motion},}\ }\href@noop {} {\bibfield  {journal} {\bibinfo
  {journal} {Proc. Natl. Acad. Sci. U.S.A.}\ }\textbf {\bibinfo {volume}
  {92}},\ \bibinfo {pages} {6705} (\bibinfo {year} {1995})}\BibitemShut
  {NoStop}%
\bibitem [{\citenamefont {Besson}\ \emph {et~al.}(1996)\citenamefont {Besson},
  \citenamefont {Edwards},\ and\ \citenamefont {Tuckerman}}]{Besson1996}%
  \BibitemOpen
  \bibfield  {author} {\bibinfo {author} {\bibfnamefont {T.}~\bibnamefont
  {Besson}}, \bibinfo {author} {\bibfnamefont {W.~S.}\ \bibnamefont {Edwards}},
  \ and\ \bibinfo {author} {\bibfnamefont {L.~S.}\ \bibnamefont {Tuckerman}},\
  }\bibfield  {title} {\enquote {\bibinfo {title} {Two-frequency parametric
  excitation of surface waves},}\ }\href@noop {} {\bibfield  {journal}
  {\bibinfo  {journal} {Phys. Rev. E}\ }\textbf {\bibinfo {volume} {54}},\
  \bibinfo {pages} {507} (\bibinfo {year} {1996})}\BibitemShut {NoStop}%
\bibitem [{\citenamefont {Zhang}\ and\ \citenamefont
  {Vi{\~n}als}(1997)}]{zhang1997pattern}%
  \BibitemOpen
  \bibfield  {author} {\bibinfo {author} {\bibfnamefont {W.}~\bibnamefont
  {Zhang}}\ and\ \bibinfo {author} {\bibfnamefont {J.}~\bibnamefont
  {Vi{\~n}als}},\ }\bibfield  {title} {\enquote {\bibinfo {title} {Pattern
  formation in weakly damped parametric surface waves},}\ }\href@noop {}
  {\bibfield  {journal} {\bibinfo  {journal} {J. Fluid Mech.}\ }\textbf
  {\bibinfo {volume} {336}},\ \bibinfo {pages} {301} (\bibinfo {year}
  {1997})}\BibitemShut {NoStop}%
\bibitem [{\citenamefont {Kudrolli}\ \emph {et~al.}(1998)\citenamefont
  {Kudrolli}, \citenamefont {Pier},\ and\ \citenamefont
  {Gollub}}]{kudrolli1998superlattice}%
  \BibitemOpen
  \bibfield  {author} {\bibinfo {author} {\bibfnamefont {A.}~\bibnamefont
  {Kudrolli}}, \bibinfo {author} {\bibfnamefont {B.}~\bibnamefont {Pier}}, \
  and\ \bibinfo {author} {\bibfnamefont {J.~P.}\ \bibnamefont {Gollub}},\
  }\bibfield  {title} {\enquote {\bibinfo {title} {Superlattice patterns in
  surface waves},}\ }\href@noop {} {\bibfield  {journal} {\bibinfo  {journal}
  {Physica D}\ }\textbf {\bibinfo {volume} {123}},\ \bibinfo {pages} {99}
  (\bibinfo {year} {1998})}\BibitemShut {NoStop}%
\bibitem [{\citenamefont {Silber}\ \emph {et~al.}(2000)\citenamefont {Silber},
  \citenamefont {Topaz},\ and\ \citenamefont {Skeldon}}]{silber2000two}%
  \BibitemOpen
  \bibfield  {author} {\bibinfo {author} {\bibfnamefont {M.}~\bibnamefont
  {Silber}}, \bibinfo {author} {\bibfnamefont {C.~M.}\ \bibnamefont {Topaz}}, \
  and\ \bibinfo {author} {\bibfnamefont {A.~C.}\ \bibnamefont {Skeldon}},\
  }\bibfield  {title} {\enquote {\bibinfo {title} {Two-frequency forced
  {F}araday waves: weakly damped modes and pattern selection},}\ }\href@noop {}
  {\bibfield  {journal} {\bibinfo  {journal} {Physica D}\ }\textbf {\bibinfo
  {volume} {143}},\ \bibinfo {pages} {205} (\bibinfo {year}
  {2000})}\BibitemShut {NoStop}%
\bibitem [{\citenamefont {Arbell}\ and\ \citenamefont
  {Fineberg}(2002)}]{arbell2002pattern}%
  \BibitemOpen
  \bibfield  {author} {\bibinfo {author} {\bibfnamefont {H.}~\bibnamefont
  {Arbell}}\ and\ \bibinfo {author} {\bibfnamefont {J.}~\bibnamefont
  {Fineberg}},\ }\bibfield  {title} {\enquote {\bibinfo {title} {Pattern
  formation in two-frequency forced parametric waves},}\ }\href@noop {}
  {\bibfield  {journal} {\bibinfo  {journal} {Phys. Rev. E}\ }\textbf {\bibinfo
  {volume} {65}},\ \bibinfo {pages} {036224} (\bibinfo {year}
  {2002})}\BibitemShut {NoStop}%
\bibitem [{\citenamefont {Porter}\ \emph {et~al.}(2004)\citenamefont {Porter},
  \citenamefont {Topaz},\ and\ \citenamefont {Silber}}]{porter2004pattern}%
  \BibitemOpen
  \bibfield  {author} {\bibinfo {author} {\bibfnamefont {J.}~\bibnamefont
  {Porter}}, \bibinfo {author} {\bibfnamefont {C.~M.}\ \bibnamefont {Topaz}}, \
  and\ \bibinfo {author} {\bibfnamefont {M.}~\bibnamefont {Silber}},\
  }\bibfield  {title} {\enquote {\bibinfo {title} {Pattern control via
  multifrequency parametric forcing},}\ }\href@noop {} {\bibfield  {journal}
  {\bibinfo  {journal} {Phys. Rev. Lett}\ }\textbf {\bibinfo {volume} {93}},\
  \bibinfo {pages} {034502} (\bibinfo {year} {2004})}\BibitemShut {NoStop}%
\bibitem [{\citenamefont {Porter}\ and\ \citenamefont
  {Silber}(2004)}]{porter2004resonant}%
  \BibitemOpen
  \bibfield  {author} {\bibinfo {author} {\bibfnamefont {J.}~\bibnamefont
  {Porter}}\ and\ \bibinfo {author} {\bibfnamefont {M.}~\bibnamefont
  {Silber}},\ }\bibfield  {title} {\enquote {\bibinfo {title} {Resonant triad
  dynamics in weakly damped {F}araday waves with two-frequency forcing},}\
  }\href@noop {} {\bibfield  {journal} {\bibinfo  {journal} {Physica D}\
  }\textbf {\bibinfo {volume} {190}},\ \bibinfo {pages} {93} (\bibinfo {year}
  {2004})}\BibitemShut {NoStop}%
\bibitem [{\citenamefont {Ding}\ and\ \citenamefont
  {Umbanhowar}(2006)}]{ding2006enhanced}%
  \BibitemOpen
  \bibfield  {author} {\bibinfo {author} {\bibfnamefont {Y.}~\bibnamefont
  {Ding}}\ and\ \bibinfo {author} {\bibfnamefont {P.}~\bibnamefont
  {Umbanhowar}},\ }\bibfield  {title} {\enquote {\bibinfo {title} {Enhanced
  {F}araday pattern stability with three-frequency driving},}\ }\href@noop {}
  {\bibfield  {journal} {\bibinfo  {journal} {Phys. Rev. E}\ }\textbf {\bibinfo
  {volume} {73}},\ \bibinfo {pages} {046305} (\bibinfo {year}
  {2006})}\BibitemShut {NoStop}%
\bibitem [{\citenamefont {Skeldon}\ and\ \citenamefont
  {Guidoboni}(2007)}]{skeldon2007pattern}%
  \BibitemOpen
  \bibfield  {author} {\bibinfo {author} {\bibfnamefont {A.~C.}\ \bibnamefont
  {Skeldon}}\ and\ \bibinfo {author} {\bibfnamefont {G.}~\bibnamefont
  {Guidoboni}},\ }\bibfield  {title} {\enquote {\bibinfo {title} {Pattern
  selection for {F}araday waves in an incompressible viscous fluid},}\
  }\href@noop {} {\bibfield  {journal} {\bibinfo  {journal} {SIAM J. Appl.
  Math.}\ }\textbf {\bibinfo {volume} {67}},\ \bibinfo {pages} {1064} (\bibinfo
  {year} {2007})}\BibitemShut {NoStop}%
\bibitem [{\citenamefont {Rucklidge}\ and\ \citenamefont
  {Silber}(2009)}]{rucklidge2009design}%
  \BibitemOpen
  \bibfield  {author} {\bibinfo {author} {\bibfnamefont {A.~M.}\ \bibnamefont
  {Rucklidge}}\ and\ \bibinfo {author} {\bibfnamefont {M.}~\bibnamefont
  {Silber}},\ }\bibfield  {title} {\enquote {\bibinfo {title} {Design of
  parametrically forced patterns and quasipatterns},}\ }\href@noop {}
  {\bibfield  {journal} {\bibinfo  {journal} {SIAM J. Appl. Dyn. Syst.}\
  }\textbf {\bibinfo {volume} {8}},\ \bibinfo {pages} {298} (\bibinfo {year}
  {2009})}\BibitemShut {NoStop}%
\bibitem [{\citenamefont {Rucklidge}\ \emph {et~al.}(2012)\citenamefont
  {Rucklidge}, \citenamefont {Silber},\ and\ \citenamefont
  {Skeldon}}]{rucklidge2012three}%
  \BibitemOpen
  \bibfield  {author} {\bibinfo {author} {\bibfnamefont {A.~M.}\ \bibnamefont
  {Rucklidge}}, \bibinfo {author} {\bibfnamefont {M.}~\bibnamefont {Silber}}, \
  and\ \bibinfo {author} {\bibfnamefont {A.~C.}\ \bibnamefont {Skeldon}},\
  }\bibfield  {title} {\enquote {\bibinfo {title} {Three-wave interactions and
  spatiotemporal chaos},}\ }\href@noop {} {\bibfield  {journal} {\bibinfo
  {journal} {Phys. Rev. Lett.}\ }\textbf {\bibinfo {volume} {108}},\ \bibinfo
  {pages} {074504} (\bibinfo {year} {2012})}\BibitemShut {NoStop}%
\bibitem [{\citenamefont {Skeldon}\ and\ \citenamefont
  {Rucklidge}(2015)}]{skeldon2015can}%
  \BibitemOpen
  \bibfield  {author} {\bibinfo {author} {\bibfnamefont {A.~C.}\ \bibnamefont
  {Skeldon}}\ and\ \bibinfo {author} {\bibfnamefont {A.~M.}\ \bibnamefont
  {Rucklidge}},\ }\bibfield  {title} {\enquote {\bibinfo {title} {Can weakly
  nonlinear theory explain {F}araday wave patterns near onset?}}\ }\href@noop
  {} {\bibfield  {journal} {\bibinfo  {journal} {J. Fluid Mech.}\ }\textbf
  {\bibinfo {volume} {777}},\ \bibinfo {pages} {604} (\bibinfo {year}
  {2015})}\BibitemShut {NoStop}%
\end{thebibliography}

%

\end{document}